\documentclass[12pt]{iopart}
\usepackage{fancyhdr}
\usepackage{graphicx}

\usepackage{iopams}
\usepackage{lscape}
\usepackage{multirow}
\usepackage[colorlinks=true, citecolor=blue, linkcolor=blue]{hyperref}
\usepackage{cite}

\begin{document}
\title[Rigorous Calculations of Non-Abelian Statistics in the Kitaev Honeycomb Model]{Rigorous Calculations of Non-Abelian Statistics in the Kitaev Honeycomb Model}
\author{Ahmet Tuna Bolukbasi$^{1}$, Jiri Vala$^{1,2}$}
\address{$ˆ1$ Department of Mathematical Physics, National University of Ireland, Maynooth, Ireland \\ $^2~$School of Theoretical Physics, Dublin Institute for Advanced Studies, 10 Burlington Road, Dublin 4, Ireland}
\eads{\mailto{jiri.vala@nuim.ie}}

\begin{abstract}
We develop a rigorous and highly accurate technique for calculation of the Berry phase in systems with a quadratic Hamiltonian within the context of the Kitaev honeycomb lattice model. The method is based on the recently found solution of the model which uses the Jordan-Wigner-type fermionization in an exact effective spin-hardcore boson representation. We specifically simulate the braiding of two non-Abelian vortices (anyons) in a four vortex system characterized by a two-fold degenerate ground state. The result of the braiding is the non-Abelian Berry matrix which is in  excellent agreement with the predictions of the effective field theory. The most precise results of our simulation are characterized by an error on the order of $10^{-5}$ or lower. We observe exponential decay of the error with the distance between vortices, studied in the range from one to nine plaquettes. We also study its correlation with the involved energy gaps and provide preliminary analysis of the relevant adiabaticity conditions. The work allows to investigate the Berry phase in other lattice models including the Yao-Kivelson model and particularly the square-octagon model. It also opens the possibility of studying the Berry phase under non-adiabatic and other effects which may constitute important sources of errors in topological quantum computation. 
\end{abstract}

\maketitle

\section{Introduction}\label{sec1}
Quantum statistics is intimately related to the adiabatic exchange of indistinguishable particles. Exchanging two particles twice results in a loop trajectory which in three dimensional space can be smoothly contracted to a point, equivalent to no trajectory. The particles' wavefunction thus remains unchanged after two subsequent exchanges, and after one exchange can transform either in a symmetric or an antisymmetric fashion, giving  Bose-Einstein and Fermi-Dirac statistics respectively. In two dimensional space such a contraction of the particles' trajectory is impossible. This gives rise to a different type of quantum statistics. Algebraically, adiabatic exchange operators correspond to the elements of the permutation group $S_{N}$ for three and higher dimensional systems, and to the elements of the braid group $B_{N}$ for two dimensional systems \cite{Laidlaw1971, Wu1984}. Both groups are formed by $N-1$ generators $\tau _{1},...,\tau _{N-1}$, obeying the constraints
\begin{eqnarray}
&&\tau _{i}\tau _{j}=\tau _{j}\tau _{i},\textrm{ \ \ \ \ }|i-j|\geq 2,
\label{braid1} \\
&&\tau _{i}\tau _{i+1}\tau _{i}=\tau _{i+1}\tau _{i}\tau _{i+1} \label{braid2}\\
&&\tau _{i}^{2}=1, \textrm{\ \ \ \  (\textit{only for} $S_{N}$)} \label{braid3}
\end{eqnarray}%
where the generator $\tau _{i}$ interchanges the two particles at the positions $i$ and $i+1$. The quantum statistics then arises from the unitary irreducible representations (irreps) of these groups \cite{Laidlaw1971}. The group $S_{N}$ possesses two one-dimensional irreps which correspond to bosonic and fermionic statistics. Its higher dimensional irreps can be replaced by bosonic and fermionic statistics when a hidden degree of freedom is introduced \cite{Druhl1970}.

On the other hand, one-dimensional representations of the braid group $B_{N}$ can be any phase $e^{i\theta }$ where $\theta \in \lbrack 0,2\pi )$ (hence the name \textit{anyonic}) \cite{Wilczek1984, Leinaas1977}. More interestingly, the braid group also permits multi-dimensional unitary irreducible representations which give rise to non-Abelian statistics. Any exchange of particles then leads to a unitary rotation of a state vector of the system within a $D$-fold degenerate ground state. The degree of degeneracy $D$ depends only on the presence of $N$ well-separated identical particles. As this is solely linked to the topology of the underlying configuration space, braiding the particles is the only way to induce nontrivial operations within this ground state subspace. Consequently, the system is immune to any local perturbations or fluctuations as long as these do not exceed the spectral gap which separates the rest of the system's spectrum from this decoherence free subspace \cite{Nayak2008}. This capability to implement unitary operations within an intrinsically fault-tolerant framework offers promising applications in quantum information processing, specifically topological quantum computation \cite{Nayak2008, Freedman2002}.

A larger body of research results suggests that non-Abelian anyons are physically realized as localized quasiparticle excitations of many-body systems. These for example include the systems that manifest the fractional quantum Hall (FQH) effect \cite{Moore1991, Nayak1996}, $p_{x}+ip_{y}$ superconductors \cite{Read2000} and the Kitaev honeycomb spin lattice model \cite{Kitaev2006} (with its proposed realizations \cite{Duan_03,micheli2006}). Theoretical studies show that all these systems can be effectively described by a similar topological quantum field theory (the Ising and the related $SU(2)_2$ Chern-Simons theory) which can be characterized by three particle types labeled as $\textbf{1}$ (i.e. vacuum or trivial topological charge), $\epsilon$ (fermion) and $\sigma$ (the non-Abelian anyon). These satisfy certain fusion and braiding rules which will be specified later. The experimental study of non-Abelian anyons is indeed of fundamental importance and so far the experimental observations have yielded encouraging results, though it is to be said that final verification of non-Abelian statistics remains a great challenge \cite{Stern2010}.

Physically, anyonic statistics arise from the evolution of the system under adiabatic interchanges of these quasiparticles. According to the adiabatic approximation, a physical system remains in its instantaneous eigenspace if a given perturbation is acting on it slowly enough and if there is a gap between the corresponding eigenvalue and the rest of the Hamiltonian's spectrum.  When these perturbations draw a smooth and closed trajectory $\mathbf{C}(\lambda)$ in the parameter space, the unitary evolution of the system in the $n$ dimensional eigenspace is given by the Berry phase (matrix) $\mathcal{B}(\mathbf{C})$
\begin{equation}
\mathcal{B}(\mathbf{C}):=\mathcal{P}\exp \left\{ i\oint \mathcal{A}(%
\lambda)d\lambda\right\}
\end{equation}%
where $\mathcal{P}$ denotes the path-ordered integral and
\begin{equation*}
\mathcal{A}_{kl}(\lambda)=i\langle \Phi ^{k}(\lambda)|\frac{d}{d\lambda%
}\Phi ^{l}(\lambda)\rangle\textrm{ \ \ \ \ }k,l=\{1,...,n\}.
\end{equation*}
where $|\Phi ^{l}(\lambda)\rangle$ and $|\Phi ^{k}(\lambda)\rangle$ are the eigenstates of the system's Hamiltonian at the value of the parameter $\lambda$.

In this paper, we directly evaluate the non-Abelian statistics of the Ising anyons of the Kitaev honeycomb model. In particular, we numerically calculate the non-Abelian Berry phase (matrix) which governs the evolution of the system under the adiabatic exchange of two $\sigma$-particles (vortices) of the Kitaev honeycomb model. This work can be seen as an accurate test of the non-Abelian statistics in the Kitaev model which offers applications in the context of topological quantum information processing and computation. Moreover, it provides a direct way to study the non-Abelian statistics in lattice models with a quadratic fermionic Hamiltonian and as such it complements similar efforts carried out in the context of continuous systems \cite{Arovas_84,Baraban_09,Cheng_09}.

The non-Abelian Berry phase calculations in the Kitaev model have been a subject of study by Lahtinen and Pachos \cite{Lahtinen2009}. They developed an interesting technique for inducing the vortex motion in the Kitaev honeycomb lattice model which we have utilized in the present work, though within a different solution of the model. While the previous studies established the non-Abelian nature of the statistics, the results on both the exact form of the braid matrix and on the exponential convergence with vortex separation were not conclusive. 

We establish these results rigorously for much larger vortex separations and to a very high accuracy, but we would like to emphasize that our approach goes beyond a mere technical improvement. Our calculations rely on the solution of the Kitaev model which was presented recently by Kells et al. \cite{Kells2009}. This solution employs the Jordan-Wigner-type fermionization in the exact effective spin-hardcore boson representation of the model and uses no redundant degrees of freedom, thus allowing us to work directly with physical eigenstates of the system. This allows us to calculate the Berry phase associated with braiding vortices at the minimal distances for up to nine plaquettes. The simulation also reaches a very high degree of accuracy as measured by the Frobenius distance between the Berry matrix obtained from our simulation and the exact Berry phase from the effective field theory. The accuracy of our calculations increases exponentially fast with the vortex distance, achieving results which are characterized by errors on the order of $10^{-5}$ and lower. 

We thus present in this work a very accurate technique for the non-Abelian Berry phase calculation. The accuracy of the calculations allows us to see the exact dependence of the simulated Berry phase on the details of the model, like the splitting of the ground state level which is intrinsic to any finite system. It also shows the dependence on an exact implementation of the braiding operations, potentially allowing for an analysis of non-adiabatic effects, similar to that provided quite recently in a different context in \cite{Cheng_11}. The importance of these effects derives from the fact that they will likely be a crucial source of errors for any topological quantum information processing and computation. Possible applications thus extend to modeling and implementation of  quantum information protocols whose reliability can be tested under various effects; for example, disorder. Naturally, the first step in this potentially fruitful story is the demonstration of highly accurate and sufficiently large scale direct simulation of the Berry phase  in the Kitaev model, as presented in this manuscript.

The results we present show strong agreement with the statistical properties of Ising anyons derived from the effective theory. One can naturally wonder what the meaning of calculating the Berry phase is if we already know it from the relevant effective field theory. The point here is that the effective theory gives us nearly no ground to test the stability of the Berry phase under the effects mentioned above or to make any conclusions about its implementation under (more) realistic conditions and about what pitfalls we should expect in such situations. Moreover, the accurate calculations like those presented here provide important predictive power in the analysis of topological phases in other lattice models including, for example, the Yao-Kivelson model  \cite{Yao2007}  and the square-octagon model \cite{Kells2010c} which exhibits a kaleidoscope of topological phases including the Ising and $SU(2)_2$ phases. We believe that this is highly relevant to implementation of quantum information processing in Majorana fermion systems \cite{Alicea2010}.

The paper is organized as follows. In Section \ref{Sec_Kitaev}, we review the exact solution of the Kitaev model on a honeycomb lattice by explicitly describing its eigenstates based on the method presented in \cite{Kells2009}. We first define Jordan-Wigner types fermions in the honeycomb lattice and then represent the original Hamiltonian using these fermions. The resulting Hamiltonian is in a quadratic fermionic form which can be solved exactly. Then in Section \ref{Sec_Adia}, we discuss how to smoothly move the vortices within our solution of the model and investigate the adiabaticity of the anyonic motion. When the anyons follow a cyclic path adiabatically, the evolution of the system is governed by the Berry matrix; the numerical method for calculating this matrix will be presented in Section \ref{Sec_Num}. The presented method can be applied to the Berry matrix of any system having a quadratic fermionic Hamiltonian. In Section \ref{Sec_Res}, we discuss the results of the numerical calculation by comparing them with the expected statistics and present an error analysis of the numerical calculations.

\section{The Kitaev Honeycomb Model}  \label{Sec_Kitaev}

\begin{figure}[h!]
\begin{center}
\includegraphics[width=350pt]{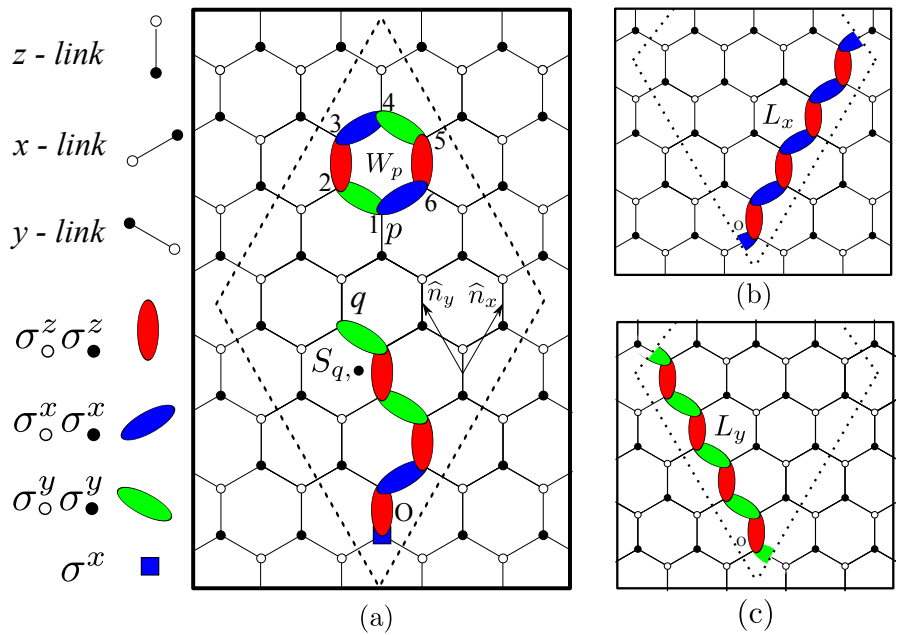}
\caption{(Color online) Kitaev honeycomb lattice model (more details in the text). Vertex
coloring emphasizes the bipartite lattice structure. The operators $W_{p}$, $S_{q,\bullet}$, $L_{x}$ and $L_{y}$ are defined as products of single- and two-body terms.}
\label{honeylattices}
\end{center}
\end{figure}

The Kitaev honeycomb model is a spin-$1/2$ lattice model in which spins are located
on the vertices of a honeycomb lattice (see Figure \ref{honeylattices}(a)), and it has the following Hamiltonian:
\begin{equation}
\mathcal{H}=-\sum_{x -\textrm{links}}J_{i,j}^{x }K_{i,j}^{x}-\sum_{y -\textrm{links}}J_{i,j}^{y }K_{i,j}^{y}-\sum_{z -\textrm{links}}J_{i,j}^{z }K_{i,j}^{z}  \label{kitaev original},
\end{equation}%
where $i$ and $j$ are the position indices of the spins, $J_{i,j}^{\alpha}$, $\alpha = x, y ,z$,  are the coupling coefficients of the two-body interaction operator $K_{i,j}^{\alpha}=\sigma _{i}^{\alpha}\sigma _{j}^{\alpha}$ on the link $(i, j)$, and the $\sigma_{i}^{\alpha}$ are the Pauli operators.

The model is exactly solvable and contains three equivalent gapped A phases for parameters satisfying $J^{x}>J^{y}+J^{z}$, $J^{y}>J^{x}+J^{z}$ or $J^{z}>J^{x}+J^{y}$, and a gapless B phase for the other values of the parameters. Furthermore, adding to the Hamiltonian \eref{kitaev original} a term which breaks the time-reversal and parity invariance of the model opens a spectral gap in the B phase and allows the realization of non-Abelian anyons of the Ising type. This additional term, defined as
\begin{eqnarray}\label{eff}
\mathcal{V}=\sum_{l=1}^{6}P_{p}^{l} &=& P_{p}^{1}+P_{p}^{2}+P_{p}^{3}+P_{p}^{4}+P_{p}^{5}+P_{p}^{6}\\
&=& \kappa _{p}^{1}~\sigma _{1}^{x}\sigma _{6}^{y}\sigma
_{5}^{z}+\kappa _{p}^{2}~\sigma _{2}^{z}\sigma _{3}^{y}\sigma
_{4}^{x}+\kappa _{p}^{3}~\sigma _{1}^{y}\sigma _{2}^{x}\sigma_{3}^{z} + \kappa _{p}^{4}~\sigma _{4}^{y}\sigma _{5}^{x}\sigma
_{6}^{z}\nonumber \\
                  &&+\kappa _{p}^{5}~\sigma _{3}^{x}\sigma _{4}^{z}\sigma
_{5}^{y}+\kappa _{p}^{6}~\sigma _{2}^{y}\sigma _{1}^{z}\sigma
_{6}^{x}\nonumber
\end{eqnarray}
where $p$ indexes honeycomb plaquettes, represents the time-reversal and parity invariance breaking effect of a weak magnetic field \cite{Kitaev2006}
\begin{equation} \label{pot}
    V= -\sum_{j}(h_{x}\sigma^{x}_{j}+h_{y}\sigma^{y}_{j}+h_{z}\sigma^{z}_{j})
\end{equation}
as it emerges from perturbation theory on the third order. The coefficients $\kappa _{p}^{l}$ of the effective term $l$ at the plaquette $p$ (Figure \ref{honeylattices}(a)) are related to the magnetic field as $\kappa \thicksim \frac{h_{x}h_{y}h_{z}}{J^{2}}$ for $J=J^{x}=J^{y}=J^{z}$. In what follows we will consider only the effective magnetic field \eref{eff}.

The model has a commuting set of plaquette operators
\begin{equation*}
W_{p}:=K_{1,2}^{y}K_{2,3}^{z}K_{3,4}^{x}K_{4,5}^{y}K_{5,6}^{z}K_{6,1}^{x}=\sigma _{1}^{z}\sigma _{2}^{x}\sigma _{3}^{y}\sigma _{4}^{z}\sigma
_{5}^{x}\sigma_{6}^{y}
\end{equation*}
for each hexagon $p$ which also commute with the total Hamiltonian $\mathcal{H}_{tot}=\mathcal{H} + \mathcal{V}$. The eigenvalues of $W_{p}$ correspond to whether the plaquette $p$ is occupied by a vortex ($-1$) or not ($+1$). The vortices carry unpaired Majorana modes for odd values of the Chern number $\nu$ \cite{Kitaev2006}. For translationally invariant configurations of the Kitaev honeycomb model, it is found that $\nu=\pm1$ depending on the direction of the magnetic field \cite{Kitaev2006}. Majorana modes exhibit non-Abelian statistics \cite{Ivanov2001} corresponding to $\sigma$-particles of the Ising anyons. We will discuss their properties in more detail in Section \ref{Sec_Res}.

The model can be solved by various fermionization techniques, but here we will use the solution introduced in the paper \cite{Kells2009} for our purposes. This solution has the advantage of giving the eigenstates of the system explicitly whereas it is not practically possible to do that in Kitaev's original solution. The original solution maps the spin degrees of freedom of the model to Majorana fermions. This requires each spin degree of freedom to be embedded in an extended Hilbert space of four dimensions. Obtaining physical states then requires projections from the eigenstates of the extended Hamiltonian which is hard to achieve in practice and thus limits the extent of numerical calculations. For the sake of self-completeness and clarity for further arguments, we start with a brief discussion of the solution we will rely on.

\subsection{The exact solution of the model}

Here we focus on $N_{x} \times N_{y}$ lattices on a torus where $N_{x}$ and $ N_{y}$ are the numbers of $z$-links in the $\widehat{n}_x$ and $\widehat{n}_y$ directions respectively, as in Figure \ref{honeylattices}(a), where the dotted lines define a 4-by-4 lattice whose opposite sides are identified. Let us label the $z$-links by $q=(q_{x},q_{y})$ with respect to the $z$-link at the origin $O=(1,1)$. Spins are denoted by either empty $\circ$ or full $\bullet$ circles, reflecting the bipartite structure of the underlying honeycomb lattice. Periodicity on the torus imposes $\prod_{p} W_{p}=1$ on the plaquette operators and gives rise to homologically nontrivial loop symmetry operators $L_{x}$ and $L_{y}$ (see Figure \ref{honeylattices}(b), (c)) which have $\pm1$ eigenvalues and commute with all $W_{p}$ and the total Hamiltonian \cite{Kells2008a}. Therefore, for a $2N$-spin system on a torus, there are $N+1$ independent operators which split the total Hilbert space of the system into $2^{N+1}$ different $2^{N-1}$-dimensional subspaces.

Before we define the fermions on the lattice, let us define the string operators between an arbitrary location on the lattice $q=(q_{x},q_{y})$ and the origin $O=(1,1)$ (see Figure \ref{honeylattices}(a)) as $S_{q,\bullet}:=S_{y}S_{x}\sigma _{q,\bullet }^{x}$ and $S_{q,\circ}:=\sigma _{q,\circ }^{z}\sigma _{q,\bullet }^{z}S_{q,\bullet}$
where the string $S_{x}$ denotes the successive applications of $\sigma _{\circ }^{z}\sigma _{\bullet }^{z}$ and $\sigma_{\bullet }^{x}\sigma _{\circ }^{x}$ to the $z$-links and $x$-links of the interval $[O$, $(q_{x},1))$ respectively and similarly $S_{y}$ is the successive applications of $\sigma_{\bullet }^{z}\sigma _{\circ }^{z}$ and $\sigma _{\circ }^{y}\sigma _{\bullet }^{y}$ to the $z$-links and $y$-links of the interval $[(q_{x},1)$, $(q_{x},q_{y}))$ respectively. Note that $S_{x}=I$ when $q_{x}=1$ and $S_{y}=I$ when $q_{y}=1$.
$S_{q,\bullet}$ and $S_{q,\circ}$ commute with $L_{y}$ and all plaquette operators except the one located on the left of the origin (i.e. $W_{(N_{x},1)}$) and $L_{x}$, with which they anti-commute.

By using string operators, fermionic creation and annihilation operators are defined on each $z$-link $q$ as
\begin{equation}
c_{q}^{\dag }=\frac{S_{q,\bullet}-S_{q,\circ}}{2},\textrm{ \ \ \ } c_{q}=\frac{S_{q,\bullet}+S_{q,\circ}}{2}
\end{equation}
It is not difficult to show that they satisfy the fermionic anti-commutation relations
\begin{equation}
\{c_{q}^{\dag },c_{q^{\prime }}\}=\delta _{q,q^{\prime }},\textrm{ \ \ \ }%
\{c_{q}^{\dag },c_{q^{\prime }}^{\dag }\}=\{c_{q},c_{q^{\prime }}\}=0.
\end{equation}
Because the $c$-fermions and the string operators have the same commutation and anti-commutation relations with the plaquette and the loop operators, the quadratic forms of fermionic operators ( e.g. $c_{q}c_{q'}^{\dag }$ for any $q$ and $q'$) commute with $L_{x}$, $L_{y}$ and $W_{q}$ for all $q$. In other words, for a $2N$-spin system on a torus, the even fermionic states span the $2^{N-1}$ dimensional subspaces of each $\left\{W_{p},L_{x},L_{y}\right\} $ configuration. Therefore, only states having an even number of $c$-fermions are realized.

\begin{figure}[t!]
\begin{center}
\includegraphics[width=300pt]{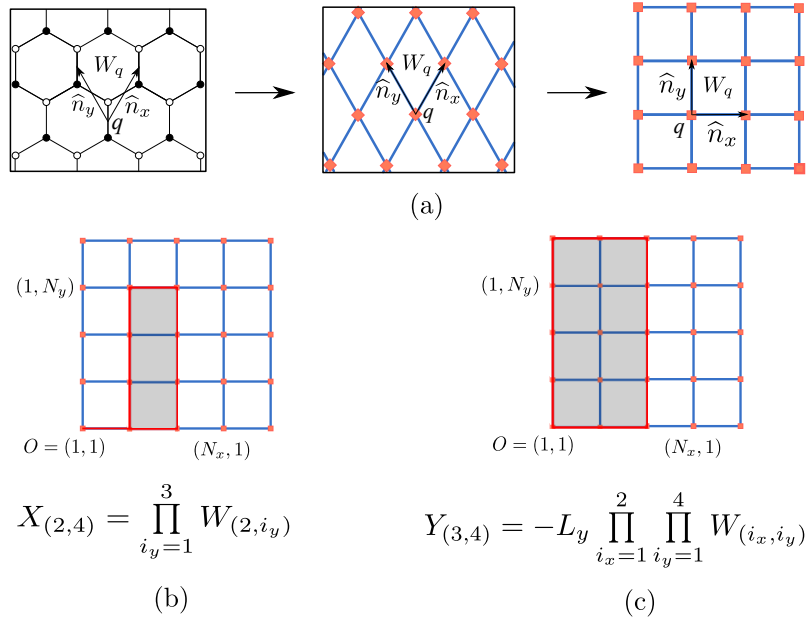}\\[0pt]
\caption{(Color online) (a) The transformation of a honeycomb lattice into a square lattice by contracting the $z$-links to a point. (b) and (c) show $X_{(2,4)}$, $Y_{(3,4)}$ respectively for a $4\times4$ square lattice whose opposite sites are identified.}\label{XY}\end{center}
\end{figure}

By using these fermions, we write $\mathcal{H}$ in Equation \eref{kitaev original} as
\begin{eqnarray}
\mathcal{H} &=&\sum_{q}J_{q}^{x}X_{q}(c_{q}^{\dag }-c_{q})(c_{q+\widehat{n}_{x}}^{\dag
}+c_{q+\widehat{n}_{x}}) \label{fermionic hamiltonian} \\
&&+\sum_{q}J_{q}^{y}Y_{q}(c_{q}^{\dag }-c_{q})(c_{q+\widehat{n}_{y}}^{\dag
}+c_{q+\widehat{n}_{y}}) \nonumber \\
\fl && +\sum_{q}J_{q}^{z}(2c_{q}^{\dag }c_{q}-I),  \nonumber
\end{eqnarray}
where $X$ and $Y$ are defined (see Figure \ref{XY}) as
\begin{eqnarray*}
X_{(q_{x},q_{y})}&=&\left\{
\begin{array}{cc}
\prod\limits_{i_{y}=1}^{q_{y}-1}W_{(q_{x},i_{y})} & \textrm{ \ \ \ \ \ \ \ }if\textrm{ \ } q_{x}\neq N_{x}  \\
-L_{x}\textrm{\ }\prod\limits_{i_{y}=1}^{q_{y}-1}W_{(q_{x},i_{y})} & \textrm{\ \ \ \ \ \ \ }if\textrm{ \ } q_{x}=N_{x}
\end{array} \right. \\
Y_{(q_{x},q_{y})}&=&\left\{
\begin{array}{cc}
1\textrm{ \ \ } & if\textrm{ \ }q_{y}\neq N_{y} \\
- L_{y}\prod\limits_{i_{x}=1}^{q_{x}-1}\prod%
\limits_{i_{y}=1}^{N_{y}}W_{(i_{x},i_{y})}\textrm{ \ \ } & if\textrm{ \ }%
q_{y}=N_{y}.%
\end{array}%
\right.
\end{eqnarray*}

Similarly, the fermionic representation of the $P_{q}^{l}$ terms of $\mathcal{V}$ \eref{eff} reads
\begin{eqnarray*}
P_{q}^{1} &=&-\kappa _{q}^{1}~iX_{q}(c_{q}^{\dag }-c_{q})(c_{q+\widehat{n}_{x}}^{\dag}-c_{q+\widehat{n}_{x}}), \\
P_{q}^{2} &=&-\kappa _{q}^{2}~iX_{q+\widehat{n}_{y}}(c_{q+\widehat{n}_{y}}^{\dag}+c_{q+\widehat{n}_{y}})(c_{q+\widehat{n}_{y}+\widehat{n}_{x}}^{\dag }+c_{q+\widehat{n}_{y}+\widehat{n}_{x}}), \\
P_{q}^{3} &=&-\kappa _{q}^{3}~iY_{q}(c_{q}^{\dag }-c_{q})(c_{q+\widehat{n}_{y}}^{\dag}-c_{q+\widehat{n}_{y}}), \\
P_{q}^{4} &=&-\kappa _{q}^{4}~iY_{q+\widehat{n}_{x}}(c_{q+\widehat{n}_{x}}^{\dag}+c_{q+\widehat{n}_{x}})(c_{q+\widehat{n}_{x}+\widehat{n}_{y}}^{\dag }+c_{q+\widehat{n}_{x}+\widehat{n}_{y}}), \\
P_{q}^{5} &=&\kappa _{q}^{5}~iX_{q+\widehat{n}_{y}}Y_{q+\widehat{n}_{x}}(c_{q+\widehat{n}_{y}}^{\dag}-c_{q+\widehat{n}_{y}})(c_{q+\widehat{n}_{x}}^{\dag }-c_{q+\widehat{n}_{x}}), \label{P5}\\
P_{q}^{6} &=&\kappa _{q}^{6}~iX_{q}Y_{q}(c_{q+\widehat{n}_{y}}^{\dag}+c_{q+\widehat{n}_{y}})(c_{q+\widehat{n}_{x}}^{\dag }+c_{q+\widehat{n}_{x}}). \label{P6}
\end{eqnarray*}%
Let us define $P_{q}^{x}$ and $P_{q}^{y}$ as
\begin{eqnarray*}
 P_{q}^{x} &:=&P_{q}^{1}+P_{q-\widehat{n}_{y}}^{2}= -\kappa _{q}^{x}~i2X_{q}(c_{q}^{\dag }c_{q+\widehat{n}_{x}}^{\dag}+c_{q}c_{q+\widehat{n}_{x}}) \\
 P_{q}^{y} &:=& P_{q}^{3}+P_{q-\widehat{n}_{x}}^{4}=-\kappa _{q}^{y}~i2Y_{q}(c_{q}^{\dag }c_{q+\widehat{n}_{y}}^{\dag}+c_{q}c_{q+\widehat{n}_{y}})
\end{eqnarray*}
where we assume $\kappa _{q}^{x}=\kappa _{q}^{1}=\kappa_{q-\widehat{n}_{y}}^{2}$, $\kappa _{q}^{y}=\kappa _{q}^{3}=\kappa_{q-\widehat{n}_{x}}^{4}$. Now, we can write the $\mathcal{V}$ term as
\begin{equation}\label{V}
\mathcal{V}=-\sum_{q}(P_{q}^{x}+P_{q}^{y}+P_{q}^{5}+P_{q}^{6}).
\end{equation}

Notice that the total Hamiltonian $\mathcal{H}_{tot} = \mathcal{H} + \mathcal{V}$ of the Kitaev model is quadratic fermionic. The quadratic fermionic Hamiltonian $H$ of any system with $N$ fermions must be of the form
\begin{equation*}
H=\frac{1}{2}\sum_{jk}\left( \xi _{jk}c_{j}^{\dag }c_{k}-\xi _{jk}^{\ast}c_{j}c_{k}^{\dag }+\Delta _{jk}c_{j}c_{k}-\Delta _{jk}^{\ast }c_{j}^{\dag}c_{k}^{\dag }\right),
\end{equation*}%
where $\xi $ is Hermitian and $\Delta $ is antisymmetric, and can be rewritten in the following way \cite{Blaizot1986}
\begin{equation}
H=\frac{1}{2}\left[
    \begin{array}{cc}
      c_{\leftrightarrow }^{\dag }&c_{\leftrightarrow }\\
    \end{array}
  \right]
  \left[
\begin{array}{cc}
\xi & \Delta \\
-\Delta ^{\ast } & -\xi ^{\ast }
\end{array}
\right]
\left[
  \begin{array}{c}
    c_{\updownarrow } \\
    c_{\updownarrow }^{\dag } \\
  \end{array}
\right]
\label{hamiltonian with M}
\end{equation}
where
\begin{eqnarray*}
  \left[
    \begin{array}{cc}
      c_{\leftrightarrow }^{\dag }&c_{\leftrightarrow }\\
    \end{array}
  \right]
 &:=& \left[c_{1}^{\dag }~...c_{_{i}}^{\dag }...~c_{N}^{\dag}~c_{1}~...c_{_{i}}...~c_{N}\right] \\
\left[
  \begin{array}{c}
    c_{\updownarrow } \\
    c_{\updownarrow }^{\dag } \\
  \end{array}
\right] &:=& \left[ c_{1}~...c_{i}...~c_{N}~c_{1}^{\dag }~...c_{_{i}}^{\dag}...~c_{N}^{\dag }\right]^{T}.
\end{eqnarray*}
Note that
\begin{eqnarray}\label{M}
M=\left[
\begin{array}{cc}
\xi & \Delta \\
-\Delta ^{\ast } & -\xi ^{\ast }
\end{array}\right]
\end{eqnarray}
is a Hermitian matrix that can be diagonalized as $M_{D}=T^{\dag }MT$ where $T$ is a unitary operator of the form
\begin{equation}\label{T}
T=\left[\begin{array}{cc}
U & V^{\ast } \\
V & U^{\ast }%
 \end{array} \right]
\end{equation}
whose columns correspond to eigenvectors of $M$. The matrix $ M_{D}:=
\left[\begin{array}{cc}
E & 0 \\
0 & -E \\
\end{array} \right]$ where $E$ is a diagonal matrix with positive entries. These are placed in an
increasing order $E_{1}<...<E_{N}$ as a convention \cite{Blaizot1986}. By replacing $M$ in Equation \eref{hamiltonian with M} with $M=TM_{D}T^{\dag }$, we get
\begin{equation*}
H=\frac{1}{2}\left[\begin{array}{cc}
      c_{\leftrightarrow }^{\dag }&c_{\leftrightarrow }\\
    \end{array} \right]
\left[\begin{array}{cc}
U & V^{\ast } \\
V & U^{\ast } \\
 \end{array} \right]%
\left[\begin{array}{cc}
E &  0\\
0 & -E%
 \end{array} \right]%
\left[\begin{array}{cc}
U^{\dag } & V^{\dag } \\
V^{T} & U^{T} \\
 \end{array} \right]%
\left[\begin{array}{cc}
c_{\updownarrow } \\
c_{\updownarrow }^{\dag } \\
 \end{array} \right].%
\end{equation*}%
Since $T$ is a unitary matrix, it is possible to define new sets of fermions
\begin{eqnarray}\label{gamma}
\left[\begin{array}{cc}
\beta _{\leftrightarrow }^{\dag } & \beta _{\leftrightarrow }\\
 \end{array} \right]
&:=&%
\left[\begin{array}{cc}
c_{\leftrightarrow }^{\dag } & c_{\leftrightarrow }\\
 \end{array} \right]%
~%
\left[\begin{array}{cc}
U & V^{\ast } \\
V & U^{\ast } \\
 \end{array} \right], \\
\left[\begin{array}{cc}
\beta _{\updownarrow } \\
\beta _{\updownarrow }^{\dag } \\
\end{array} \right]%
&:=&%
\left[\begin{array}{cc}
U^{\dag } & V^{\dag } \\
V^{T} & U^{T} \\
\end{array} \right]
\left[\begin{array}{cc}
c_{\updownarrow } \\
c_{\updownarrow }^{\dag }\\
\end{array} \right]. \nonumber
\end{eqnarray}
These two definitions are compatible, and therefore, the Hamiltonian $H$ can be rewritten in a free fermionic form as follows:
\begin{equation}
H=\sum_{i}E_{i}\beta _{i}^{\dag }\beta _{i}-\sum_{i}\frac{E_{i}}{2}.
\end{equation}

For the total Hamiltonian $\mathcal{H}_{tot}$ of the Kitaev model, $\mathcal{M}_{tot}$ -- the analog of the matrix (\ref{M}) -- is given in terms of $X$ and $Y$; therefore it is diagonalized separately for each $X$ and $Y$ configuration (i.e. $\left\{W_{p},L_{x},L_{y}\right\} $ configuration).

On the other hand, the eigenstates of the quadratic fermionic Hamiltonians can be explicitly written by using the Bloch-Messiah theorem \cite{Bloch1962, Ring2004}. According to this theorem, any
unitary matrix of the form of $T$ in Equation \eref{T} can be decomposed into three
matrices of very special form:%
\begin{equation}\label{BM1}
T=%
\left[\begin{array}{cc}
D & 0 \\
0 & D^{\ast }%
\end{array} \right]%
\left[\begin{array}{cc}
\overline{U} & \overline{V} \\
\overline{V} & \overline{U}%
\end{array} \right]%
\left[\begin{array}{cc}
C & 0 \\
0 & C^{\ast }%
\end{array} \right]%
,
\end{equation}%
where $D$ and $C$ are unitary matrices and $\overline{U}$ and $\overline{V}$
are real matrices of the general block-diagonal form
\begin{equation*}
\overline{U}=\left[
  \begin{array}{ccccc}
    Z &  &  &  &  \\
     & \overline{U}_{1} &  & \mbox{\huge $0$} &  \\
     &  & \ddots &  &  \\
     & \mbox{\huge $0$}  &  & \overline{U}_{n} &  \\
     &  &  &  & I \\
  \end{array}
\right],\textrm{ \ }\overline{V}=\left[
  \begin{array}{ccccc}
    I &  &  &  &  \\
     & \overline{V}_{1} &  &\mbox{\huge $0$}   &  \\
     &  & \ddots &  &  \\
     & \mbox{\huge $0$}  &  & \overline{V}_{n} &  \\
     &  &  &  & Z \\
  \end{array}
\right]
\end{equation*}
where $Z$ and $I$ are the zero and identity matrices of the same size respectively, and
\begin{equation}\label{UV}
    \overline{U}_{i}=\left[
       \begin{array}{cc}
         u_{i} &  0\\
        0  & u_{i} \\
       \end{array}
     \right],\textrm{ \ \ \ \ \ \ \ \ \ }\overline{V}_{i}=\left[
       \begin{array}{cc}
         0 & v_{i} \\
         -v_{i} & 0\\
       \end{array}
     \right]
\end{equation}
where $u_{i}$ and $v_{i}$ are positive real numbers.
By using Equation \eref{BM1} in Equation \eref{gamma}, we have
\begin{equation*}
\left[\begin{array}{cc}  \beta _{\leftrightarrow }^{\dag }~\beta _{\leftrightarrow }\end{array} \right]=
\left[\begin{array}{cc} c_{\leftrightarrow }^{\dag }~c_{\leftrightarrow }\end{array} \right]~%
\left[\begin{array}{cc}
D & 0 \\
0 & D^{\ast }%
\end{array} \right]%
\left[\begin{array}{cc}
\overline{U} & \overline{V} \\
\overline{V} & \overline{U}%
\end{array} \right]%
\left[\begin{array}{cc}
C & 0 \\
0 & C^{\ast }%
\end{array} \right]%
.
\end{equation*}%
Here, $D$ is used to define new operators $a^{\dag }$ and $a$:
\begin{equation*}
\left[\begin{array}{cc}
a_{\leftrightarrow }^{\dag } & a_{\leftrightarrow }%
\end{array} \right]%
:=%
\left[\begin{array}{cc}
c_{\leftrightarrow }^{\dag } & c_{\leftrightarrow }%
\end{array} \right]%
~%
\left[\begin{array}{cc}
D & 0 \\
0 & D^{\ast }%
\end{array} \right]%
=\left\{
\begin{array}{c}
a_{\leftrightarrow }^{\dag }=c_{\leftrightarrow }^{\dag }D \\
\ \: a_{\leftrightarrow }=c_{\leftrightarrow }D^{\ast }%
\end{array}.
\right.
\end{equation*}%
Then there is a special Bogoliubov transformation
\begin{equation*}
\left[\begin{array}{cc}
\alpha _{\leftrightarrow }^{\dag } & \alpha _{\leftrightarrow }%
\end{array} \right]%
:=%
\left[\begin{array}{cc}
a_{\leftrightarrow }^{\dag } & a_{\leftrightarrow }%
\end{array} \right]%
~%
\left[\begin{array}{cc}
\overline{U} & \overline{V} \\
\overline{V} & \overline{U}%
\end{array} \right]%
=\left\{
\begin{array}{c}
\alpha _{\leftrightarrow }^{\dag }=a_{\leftrightarrow }^{\dag }\overline{U}%
+a_{\leftrightarrow }\overline{V} \\
\alpha _{\leftrightarrow }=a_{\leftrightarrow }^{\dag }\overline{V}%
+a_{\leftrightarrow }\overline{U}%
\end{array}%
\right. .
\end{equation*}
This distinguishes the ``paired" levels $(u_{p}>0;v_{p}>0)$%
\begin{eqnarray*}
&&\alpha _{p}^{\dag }=u_{p}a_{p}^{\dag }-v_{p}a_{\overline{p}}, \textrm{\ \ \ \ \ \ \ \ } \alpha _{p} =-v_{p}a_{\overline{p}}^{\dag }+u_{p}a_{p},\\
&&\alpha _{\overline{p}}^{\dag }=u_{p}a_{\overline{p}}^{\dag }+v_{p}a_{p}, \textrm{\ \ \ \ \ \ \ \ } \alpha _{\overline{p}} =v_{p}a_{p}^{\dag }+u_{p}a_{\overline{p}},
\end{eqnarray*}%
(where $(p,\overline{p})$ are defined by $\overline{U}_{i}$ and $\overline{V}_{i}$ \eref{UV})
from the ``blocked" levels
$(v_{m}=0;u_{m}=1)$%
\begin{eqnarray*}
\alpha _{i}^{\dag } &=&a_{i},\textrm{ \ \ \ \ \ \ \ \ }\alpha _{m}^{\dag
}=a_{m}^{\dag }, \\
\alpha _{i} &=&a_{i}^{\dag },\textrm{ \ \ \ \ \ \ \ \ }\alpha _{m}=a_{m},
\end{eqnarray*}%
which are either occupied $(v_{i}=1;u_{i}=0)$ or empty. Finally, a linear transformation of the $\alpha ^{\dag }$ and $\alpha $  by the unitary matrix $C$ gives:%
\begin{equation*}
\left[\begin{array}{cc}
\beta _{\leftrightarrow }^{\dag } & \beta _{\leftrightarrow }%
\end{array} \right]%
~:=%
\left[\begin{array}{cc}
\alpha _{\leftrightarrow }^{\dag } & \alpha _{\leftrightarrow }%
\end{array} \right]%
\left[\begin{array}{cc}
C &  0 \\
0 & C^{\ast }%
\end{array} \right]%
\Rightarrow\left\{
\begin{array}{c}
\beta _{\leftrightarrow }^{\dag }=\alpha _{\leftrightarrow }^{\dag }C \\
\ \: \beta _{\leftrightarrow }=\alpha _{\leftrightarrow }C^{\ast }%
\end{array}%
\right. .
\end{equation*}%

For a general quadratic fermionic Hamiltonian, the ground state wavefunction
is defined as a non-zero wavefunction $|$$\phi \rangle $ such that $\beta
_{k}|\phi \rangle =0$ for all $k$. It can be easily verified that the following
wavefunction satisfies these criteria:
\begin{equation}
|\phi \rangle =\prod\limits_{i}a_{i}^{\dag
}\prod\limits_{p}(u_{p}+v_{p}a_{p}^{\dag }a_{\overline{p}}^{\dag })|-\rangle
,  \label{Phi}
\end{equation}%
where $|$$-\rangle $ represents the vacuum of $c$-fermions. In the honeycomb model, the vacuum $|$$-\rangle $ belongs to the $X$,$Y$-configurations for which $\mathcal{M}_{tot}$ is diagonalized. However, because only the even fermionic configurations are allowed for each $X$,$Y$-configuration, when the number of elements in the first product (i.e. $i$-part) of \eref{Phi} is odd, the ground state $|\Phi \rangle$ of the model is the next excited state $|\Phi \rangle=\beta _{1}^{\dag }|\phi \rangle$ where $\beta _{1}^{\dag }$ is the minimum energy fermion. In that respect, the number of $a_{i}^{\dag}$ determines the fermionic parity of the system.

For future reference, it is important to associate every eigenstate of the system with a $T=\left[\begin{array}{cc}
U & V^{\ast } \\
V & U^{\ast } \\
\end{array} \right]$ matrix such that the Bloch-Messiah representation Equation \eref{Phi} represents that eigenstate. For an excited state $\beta _{i}^{\dag }|\phi \rangle$, this can be done by exchanging the roles of $\beta _{i}(i)$ and $\beta _{i}^{\dag }(i)$ which manifests itself  as the exchange of the $i$th and $(N+i)$th columns of $T$; we denote this matrix by $T'$. Thus  the ground state $|\phi' \rangle$ of $T'$ is equal to $\beta _{i}^{\dag }|\phi \rangle$. A straightforward generalization of this method to the other excited states associates any eigenstate with a particular $T$ matrix.

Ä
\section{Adiabatic Motion of Vortices and the non-Abelian Berry Phase} \label{Sec_Adia}

In this Section, we discuss how to move vortices from one plaquette to another, then describe the evolution of the system under an exchange of vortices.

By changing the sign of the relevant coupling coefficients $J$ and $\kappa$, we can emulate the fermionic spectrum relevant to any $X$,$Y$-configuration starting from the trivial $X,Y$-configuration (i.e.\ $X_{q}=1$ and $Y_{q}=1$ for all $q$) which we call the reference $X,Y$-configuration. 

This approach, first introduced in \cite{Lahtinen2009}, allows us to consider the Hilbert spaces of different $X,Y$-configurations connected by the coupling coefficients $J$ and $\kappa$.

\begin{figure}[t]
\begin{center}
\includegraphics[width=300pt]{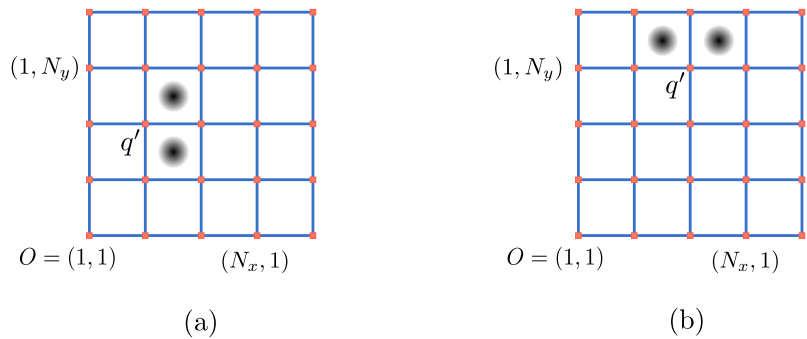}\\[0pt]
\end{center}
\caption{(Color online) (a) The creation of two vortices sharing an $x$-link. (b) The creation of two vortices sharing a $y$-link.}
\label{VortexCM}
\end{figure}

In this way, we can also simulate the creation, annihilation and motion of the vortices. For example, consider the $\left\{W_{p},L_{x},L_{y}\right\}$ configuration (see Figure \ref{VortexCM}(a)) $$L_{x}=L_{y}=-1 \textrm{ and } W_{q}=1 \textrm{ for all } q \textrm{ except } W_{q'}=W_{q'-\widehat{n}_{y}}=-1;$$ where $q'=(q' _{x},q' _{y})$ such that $q' _{y}\neq1$. This configuration gives $Y_{q}=1$ and $X_{q}=1$ for all $q$ except $X_{q'}=-1$. The fermionic spectrum of this configuration can be achieved from the reference $X,Y$-configuration by using the negative values of the set of coefficients $[J\kappa]_{q'}^{x}:=\{J_{q'}^{x},\kappa_{q'-\widehat{n}_{y}}^{5},\kappa _{q'}^{6},\kappa _{q'}^{x}\}$ (see (\ref{fermionic hamiltonian}) and (\ref{V})). In other words, changing the sign of  $[J\kappa]_{q'}^{x}:=\{J_{q'}^{x},\kappa_{q'-\widehat{n}_{y}}^{5},\kappa _{q'}^{6},\kappa _{q'}^{x}\}$ can be considered as the creation of two vortices from the vacuum (i.e. the reference $X,Y$-configuration with positive $J$ and $\kappa$ values).

For an analogous configuration shown in Figure \ref{VortexCM}(b), vortices on the plaquettes $q'$ and $q'-\widehat{n}_{x}$ for $q' _{y}=N_{y}$ and $q' _{x}\neq 1$ can be created from the reference $X,Y$-configuration by changing the sign of the set of coefficients $[J\kappa]_{q'}^{y}:=\{J_{q'}^{y},\kappa_{q'-\widehat{n}_{x}}^{5},\kappa _{q'}^{6},\kappa _{q'}^{y}\}$.

Generally, changing the sign of $[J\kappa]_{q}^{x}$ alters the vorticity of two plaquettes $W_{(q _{x},q _{y})}$ and $W_{(q _{x},q _{y}-1)}$ when $q _{y}\neq 1$, and may be seen as creation, annihilation or motion of vortices, depending on the initial vorticity of the plaquettes. A similar effect can be achieved on the plaquettes $W_{(q _{x},q _{y})}$ and $W_{(q _{x}-1,q _{y})}$ for $q$ satisfying $q _{y}=N_{y}$ and $q _{x}\neq 1$ when we change the sign of $[J\kappa]_{q}^{y}$. We point out here for completeness that it is also possible to simulate the vortices by changing the $[J\kappa]_{q}^{x}$ when $q _{y}= 1$ or by changing $[J\kappa]_{q}^{y}$ for $q$ satisfying $q_{y}\neq N_{y}$ or $q _{x}=1$ by carrying the sign to some of the $c$-fermions. However we will use the previous approach to exchange vortices as it is sufficient.

\begin{figure}[h!]
\begin{center}
\includegraphics[width=250pt]{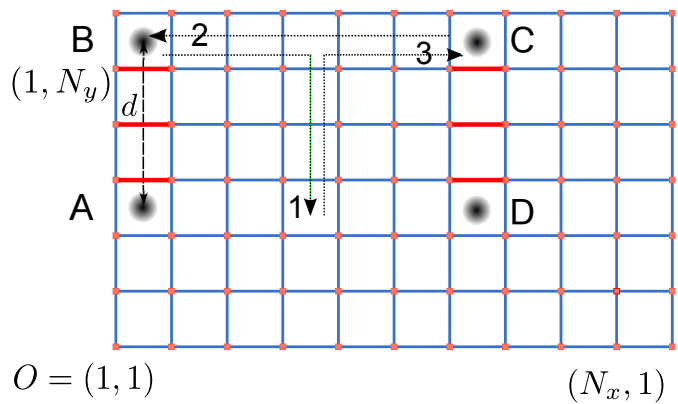}\\[0pt]
\end{center}
\caption{(Color online) A configuration with 4 vortices having minimum distance $d=3$ and sizes $N_{x}=10$, $N_{y}=6$. Opposite sides of the lattices are identified. Red links highlight the adiabatically changed links. The path swapping the vortices \textsf{\textbf{B}} and \textsf{\textbf{C}} consists of the links on the arrows 1, 2 and 3.}
\label{GConf}
\end{figure}

From now on we will use the term vortex only for these simulated vortices. In Figure \ref{GConf} a template configuration is shown with four vortices which are created from the vacuum by changing the sign of $[J\kappa]^{x}$ of the $x$-links between \textsf{\textbf{A}} and \textsf{\textbf{B}}, and between \textsf{\textbf{C}} and \textsf{\textbf{D}} (colored by red in Figure \ref{GConf}). In this paper, we will work with several configurations similar to Figure \ref{GConf} with different sizes as: $N_{x}=3d+1$, $N_{y}=2d$ when $d$ is odd and $N_{x}=3d$, $N_{y}=2d$ when $d$ is even for $d=1,...,9$, where $d$ is the minimum distance between the vortices. Note that all configurations are even-by-even; the other configurations (odd-by-odd, even-by-odd etc.) will be studied in future. For all these configurations, the coupling coefficients for the vacuum are identical for all plaquettes: $J_{q}^{x}=J_{q}^{y}=J_{q}^{z}=J=1$ for all $q$. The strength of the effective magnetic field $\kappa$ for the vacuum is also the same for all the plaquettes $q$, $\kappa_{q}^{x}=\kappa_{q}^{y}=\kappa_{q}^{5}=\kappa_{q}^{6}=\kappa$, and is taken from two sets which differ in magnitude: (i)  $\kappa=l\times0.01$ where $l=1,...,5$, and (ii) $\kappa=l\times0.05$ where $l=2,...,8$.

\begin{figure}[t]
\begin{center}
\includegraphics[width=460pt]{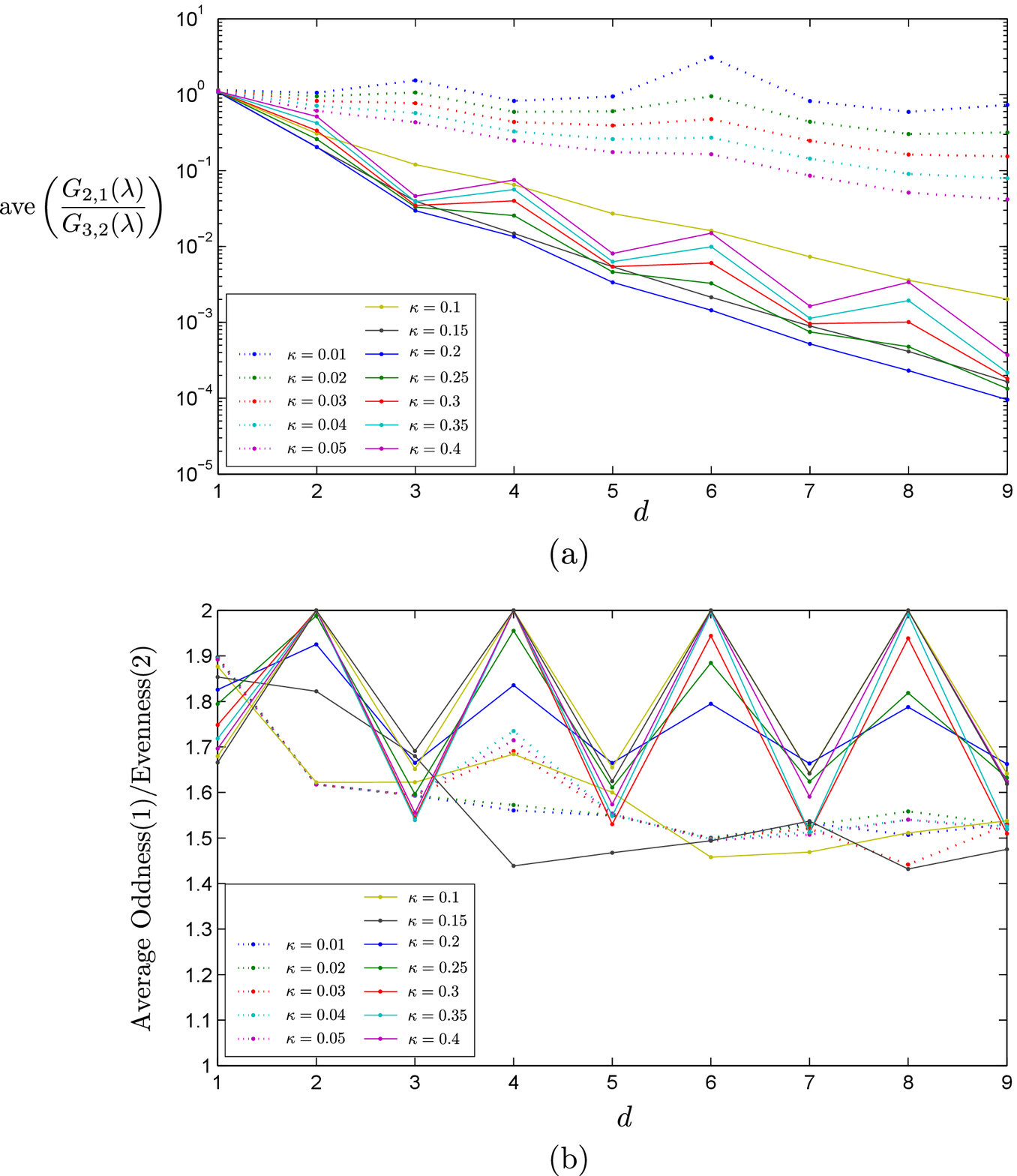}\\[0pt]
\end{center}
\caption{(Color online) (a) Average value of the ratio $G_{2,1}(\lambda)$/$G_{3,2}(\lambda)$ along the trajectory vs. minimum distance $d$ between vortices. (b) The average values of the parity of the system while the particles are exchanged vs. $d$.  Average oddness/evenness of the systems along the trajectory is calculated after assigning 2 and 1 to the even and odd parity systems respectively.  Note that the numerical details of the calculations are given at the beginning of Section \ref{Sec_Adia}.}
\label{Gap}
\end{figure}

To swap the position of the vortices \textsf{\textbf{B}} and \textsf{\textbf{C}} -- see Figure \ref{GConf} -- we need to adiabatically move the vortices along the paths indicated by the arrows 1, 2, and 3. This requires us to slowly change the sign $[J\kappa]$ of the links which are intersected by the paths. 
For the large values of $\kappa$, i.e. the values from the set (ii) above, all these configurations have a (nearly) two-fold degenerate ground state which is separated by a gap from the rest of the Hamiltonian's spectrum. This can be seen from Figure \ref{Gap}(a) which shows the average of the ratio of the splitting of the ground state degeneracy and of the gap to the first excited state $G_{2,1}(\lambda)$/$G_{3,2}(\lambda)$ along the path. Here $G_{n,m}(\lambda)=|E^{m}(\lambda)-E^{n}(\lambda)|$ is the energy difference between the $m^{th}$ and $n^{th}$ eigenstates and $\lambda$ parametrizes the path, i.e. it represents the values of $[J\kappa]$ for the links on the path. The degeneracy decreases with the minimum distance $d$. The ratio $G_{2,1}(\lambda)$/$G_{3,2}(\lambda)$ for small values of $\kappa$ (i.e. from the set (i)) exhibits more involved behavior as the gap between the ground states and higher excited levels is much smaller in this case.

It is interesting to point out that fermionic parity of the system may change while the particles are exchanged. The average values of the parity of the system are shown in Figure \ref{Gap}(b) where 2 and 1 are assigned to the even and the odd parity sectors respectively. The parity of the system sets the small gap $G_{2,1}$ between the nearly degenerate states as $G_{2,1}=E_{2}-E_{1}$ for odd systems and $G_{2,1}=E_{2}+E_{1}$ for even systems, where $E_{i}$ denotes the spectrum of the system increasing with the index $i\geq1$. However, the parity does not affect $G_{m,2}$ for any value of $m$. This is one of the causes of the oscillations seen in Figure \ref{Gap}. It is not the only cause though. Calculations show that the average energy of the nearly-zero modes ($E_{1}$ and $E_{2}$) oscillates in the same way as well.

For a system which is initially in the ground state space, the exchange of \textsf{\textbf{B}} and \textsf{\textbf{C}} transforms the system within the $2$-dimensional ground state space provided that the process is \textit{adiabatic} (\nameref{Adia_Smooth} contains a detailed discussion on adiabaticity and the path which is followed). When the change of the parameters of the Hamiltonian follow a \textit{smooth} closed curve in the parameter space, then the evolution of the system is governed by the following Berry phase (matrix) $\mathcal{B}(\mathbf{C})$ \cite{Bohm2009}:
\begin{eqnarray} \label{Berry_Kitaev}
\mathcal{B}(\mathbf{C})&:=&\mathcal{P}\exp \left\{ i\oint \mathcal{A}(%
\lambda)d\lambda\right\} \label{Berry_Kitaev1}\\
&:=&\lim_{M\rightarrow \infty }\exp \left\{ i\mathcal{A}(\lambda_{M})\Delta \lambda \right\}  \cdots \exp \left\{ i%
\mathcal{A}(\lambda_{1})\Delta \lambda \right\} \nonumber
\end{eqnarray}%
where $\mathcal{P}$ denotes the path-ordered integral,
$ \mathcal{A}_{ab}(\lambda_{k})= i \left.\langle \Phi ^{a}(\lambda)|\frac{d}{d\lambda%
}\Phi ^{b}(\lambda)\rangle\right|_{\lambda=\lambda_{k}}$, $a,b=\{1, 2\}$, and $\lambda_{1}$ and $\lambda_{M+1}$ are coinciding points denoting the beginning and end point of the closed trajectory whose curve length \textbf{length}$(\lambda_{M+1},\lambda_{1})$ is divided into $M$ equal pieces $\Delta \lambda =$\textbf{length}$(\lambda_{M+1},\lambda_{1})/M.$  Note that because the path traveled by vortices encloses zero area, all local and geometrical effects are eliminated and only the topological interaction is realized.

The Berry matrix $\mathcal{B}(\mathbf{C})$ is written in the same basis as that of $\mathcal{A}(\lambda_{1})$ (see \nameref{Adia_Smooth} for more details about $\lambda_{1}$). However, it is independent of the choice of the bases in which other $ \mathcal{A}(\lambda)$s are written, as long as the basis states are smooth functions of the parameter $\lambda$. For practical purposes, we used the nearly degenerate energy eigenstates ($|\Phi ^{1}(\lambda)\rangle$ and $|\Phi ^{2}(\lambda)\rangle$) of the system. 
We point out for clarity that $|\Phi ^{1}(\lambda)\rangle$ and $|\Phi ^{2}(\lambda\rangle$ are always distinguishable thanks to the small gap which separates them and which is never less than $10^{-6}$ for all values of $\kappa$ and $d$ we used.

\section{Numerical Methods for Calculating the non-Abelian Berry Phase} \label{Sec_Num}

In this Section, we present the arguments that we use for the numerical calculation of the Berry matrix. The results and discussions are presented in the next Section.

In order to calculate the Berry matrix $\mathcal{B}(\mathbf{C})$ (\ref{Berry_Kitaev}), we first need to evaluate
\begin{equation}\label{A}
\mathcal{A}_{ab}(\lambda_{k})= i \left.\langle \Phi ^{a}(\lambda)|\frac{d}{d\lambda}\Phi ^{b}(\lambda)\rangle\right|_{\lambda=\lambda_{k}}, \textrm{\ \ \ } a,b=\{1,2\}.
\end{equation}
There are many different ways to approximate the derivative of a function \cite{Mathews2004}, but we are going
to use the central-difference formula which says
\begin{equation}
f^{\prime }(x)\Delta x\simeq \frac{f(x+\Delta x)-f(x-\Delta x)}{2}+O(\left(
\Delta x\right) ^{3})  \label{centraldiff}
\end{equation}%
as long as the third derivative of $f$ is continuous. It approximates $f^{\prime }(x)$ on the order $O(\left(\Delta x\right) ^{2})$; however it is also possible to get higher order approximations \cite{Mathews2004}. In our case, we performed the calculations of $f^{\prime }(x)$ on the order $O(\left(\Delta x\right) ^{6})$.

By applying the formula (\ref{centraldiff}) to $\mathcal{A}_{ab}(\lambda_{k})$, we get
\begin{equation*}
\mathcal{A}_{ab}(\lambda_{k})=i\frac{\langle \Phi ^{a}(\lambda _{k})|\Phi ^{b}(\lambda _{k+1} )\rangle -\langle \Phi ^{a}(\lambda_{k})|\Phi ^{b}(\lambda_{k-1})\rangle}{2\Delta\lambda}.
\end{equation*}
This reduces the Berry matrix calculation to finding the overlaps between the ground states of adjacent points on the trajectory.

Let $M$ be the number of data points used to calculate the Berry matrix (\ref{Berry_Kitaev}). For all data points $k=1,...,M$ and some other point $k=0$, let $|\phi _{k}\rangle$ be the ground states of $\beta (k)$-fermions defined as
\begin{equation}\label{gammak}
\left[\begin{array}{cc}
\beta _{\leftrightarrow }^{\dag }(k) & \beta _{\leftrightarrow }(k) \\
 \end{array} \right]%
:=
\left[\begin{array}{cc}
c_{\leftrightarrow }^{\dag } & c_{\leftrightarrow } \\
 \end{array} \right]%
 T(k)
\end{equation}
where
$
T(k):=\left[\begin{array}{cc}
U(k) & V^{\ast }(k) \\
V(k) & U^{\ast }(k) \\
 \end{array} \right]
$ as in equation \eref{T}.
Since the $T$'s are unitary matrices, we can write
\begin{equation}\label{left}
\left[\begin{array}{cc}
c_{\leftrightarrow }^{\dag } & c_{\leftrightarrow } \\
 \end{array} \right] = \left[\begin{array}{cc}
\beta _{\leftrightarrow }^{\dag }(0) & \beta _{\leftrightarrow }(0) \\
 \end{array} \right] T^{\dag }(0)
\end{equation}
for $k=0$ and replace $\left[\begin{array}{cc}
c_{\leftrightarrow }^{\dag } & c_{\leftrightarrow } \\
 \end{array} \right]$ of Equation \eref{gammak} with the left hand side of Equation \eref{left} as
\begin{equation}\label{gammai0}
\left[\begin{array}{cc}
\beta _{\leftrightarrow }^{\dag }(k) & \beta _{\leftrightarrow }(k)\\
 \end{array} \right]=\left[\begin{array}{cc}
\beta _{\leftrightarrow }^{\dag }(0) & \beta _{\leftrightarrow }(0)\\
 \end{array} \right]T(k,0)
\end{equation}
where
\begin{equation*}
T(k,0):=T^{\dag}(0)T(k)=\left[\begin{array}{cc}
U(k,0) & V^{\ast }(k,0) \\
V(k,0) & U^{\ast }(k,0) \\
 \end{array} \right]
\end{equation*}
and
\begin{eqnarray}
U(k,0) &:=&U^{\dag }(0)\ U(k)+V^{\dag }(0)\ V(k)  \label{UV definitions} \\
V(k,0) &:=&V^{T}(0)\ U(k)+U^{T}(0)\ V(k).  \nonumber
\end{eqnarray}

By applying the Bloch-Messiah theorem to Equation \eref{gammai0}, we can only get the absolute value of the overlap between the ground states $|\langle \phi _{0}|\phi _{k}\rangle| =\sqrt{|\det U(k,0)| }$ (see \nameref{BM}). However, it is possible to get the complete overlap by using the Thouless theorem \cite{Thouless1962,Ring2004}. The Thouless theorem states that when $\langle \phi _{k}|\phi_{0}\rangle \neq 0$, the ground state $|$$\phi _{k}\rangle $ can be written as
\begin{equation}
|\phi _{k}\rangle =|\psi _{(k,0)}\rangle :=\sqrt{|\det U(k,0)| }e^{\mathbf{Z}(k,0)}|\phi _{0}\rangle \label{Thouless rep}
\end{equation}%
where
\begin{equation*}
\fl \mathbf{Z}(k,0)=\frac{1}{2}\sum_{n,n^{\prime
}}Z_{nn^{\prime }}(k,0)\beta _{n}^{\dag }(0)\beta _{n^{\prime }}^{\dag
}(0) \textrm{\ \ \ and \ \ } Z(k,0)=\left( V(k,0)U^{-1}(k,0)\right) ^{\ast }.
\end{equation*}

Having the ground state $|\phi _{k}\rangle$ represented as $|\psi _{(k,0)}\rangle$ for $k=1,...,M$ where we have fixed the overall phase to 1 (Equation \eref{Thouless rep}), we 
recall that the excited states can be represented by using the column exchange technique on $T(k,0)$ discussed at the end of Section \ref{Sec_Kitaev}.  Then, the overlap between the ground states $|\psi _{(l,0)}\rangle $ and $|\psi _{(k,0)}\rangle $ for $k\not=l\neq 0,$ reads
\begin{equation*}
\langle \psi _{(l,0)}|\psi _{(k,0)}\rangle =\sqrt{|\det U(l,0)|}\textrm{ }%
\sqrt{|\det U(k,0)|}\textrm{ \ \ }  \langle \phi _{0}|e^{\mathbf{Z}^{\dag }(l,0)}e^{%
\mathbf{Z}(k,0)}|\phi _{0}\rangle .
\end{equation*}%
In a recent paper \cite{Robledo2009}, the overlap $\langle \phi _{0}|e^{\mathbf{Z}^{\dag
}(l,0)}e^{\mathbf{Z}(k,0)}|\phi _{0}\rangle $ has been calculated as
\begin{equation*}
\langle \phi _{0}|e^{\mathbf{Z}^{\dag }(l,0)}e^{\mathbf{Z}(k,0)}|\phi
_{0}\rangle =\left( -1\right) ^{N\left( N+1\right) /2}~\textrm{Pf}(\mathcal{Z}%
(l,0;k,0))
\end{equation*}
where Pf denotes the Pfaffian, $
\mathcal{Z}(l,0;k,0)=
\left[\begin{array}{cc}
Z(k,0) & -I \\
I & -Z^{\ast }(l,0) \\
 \end{array} \right]
$ and $N$ is the number of fermions (and therefore also the size of $Z$). Since the numerical algorithms for calculating the Pfaffian of large matrices are generally slow, it is more convenient to proceed using the following expression (see \nameref{Pfaffian})
\begin{eqnarray}
\langle \psi _{(l,0)}|\psi _{(k,0)}\rangle = \left( -1\right) ^{N\left( N+1\right) /2}\sqrt{%
\exp \left\{ i\theta _{0}(l,k) \right\} ~|\det U(l,k) |}\label{overlap formula}
\end{eqnarray}%
 where
\begin{eqnarray*}
U(l,k) &:=&U^{\dag }(k)U(l)+V^{\dag }(k)V(l) \\
\theta _{0}(l,k) &=&\arg \left\{ \det U(k,0)\det U^{\dag
}(l,0)\det U(l,k) \right\}.
\end{eqnarray*}
Although the sign in Equation \eref{overlap formula} is ambiguous due to the square root, for a series of smoothly varying matrices the correct sign can be traced from their previous values.

We point out here that, for the purpose of calculating the Berry matrix \eref{Berry_Kitaev1},  the degenerate ground states at each point of the trajectory in parameter space is defined in terms of the reference state $|\phi _{0}\rangle$. Thus the reference state must be chosen in such a way that it is not orthogonal to any of the ground states along the whole trajectory, as in Equation \eref{Thouless rep}. 

\section{Numerical Results and Discussions} \label{Sec_Res}

Before presenting the numerical results, let us briefly discuss the Ising anyons consisting of three different particles: $\textbf{1}$ (i.e. vacuum), $\epsilon$ and $\sigma$. Bringing two particles together is called fusion, and Ising anyons satisfy the following fusion rules:
\begin{eqnarray}
&&\sigma \times \sigma  =\textbf{1}+\epsilon ,\textrm{ \ \ \ }\sigma \times \epsilon =\sigma ,%
\textrm{\ \ \ \ \ \ }\epsilon \times \epsilon =\textbf{1}, \label{FRules} \\
&&\textbf{1}\times \textbf{1} =\textbf{1},\textrm{ \ \  \ \ \ \ \ \ }\textbf{1}\times \sigma =\sigma ,\textrm{\ \ \ \
\ \ }\textbf{1}\times \epsilon =\epsilon \nonumber
\end{eqnarray}
where $\times$ denotes the fusion operator. For example, the first rule says that two $\sigma$-particle may either annihilate or fuse into an $\epsilon$-particle. Although the fusion of two Abelian anyons gives one outcome, non-Abelian anyons have multiple fusion possibilities or channels which are one way of accounting for the degeneracy of the ground state. Thus $\sigma$-particles are non-Abelian anyons while the other two particles are Abelian.

To detail the nontrivial implications of these rules, consider a system with four $\sigma $-particles ($a, b, c$ and $d$) whose total topological charge corresponds to the  vacuum. In this system, the fusion of any two $\sigma $-particles (say $a$ and $b$) determines the fusion channel of the other two $\sigma $-particles ($c$ and $d$) and because there are two different fusion results that $a$ and $b$ can fuse into, there is a two dimensional fusion space associated with the system. The basis of this space can be chosen as the resulting fusion states of $a$ and $b$ as $\{|\Omega^{ab}_{\footnotesize\textbf{1}}\rangle,|\Omega^{ab}_{\epsilon}\rangle\}$. On the other hand, another basis can be chosen based on the fusion results of $b$ and $c$ $\{|\Omega^{bc}_{\footnotesize\textbf{1}}\rangle,|\Omega^{bc}_{\epsilon}\rangle\}$. The matrix that relates the two fusion bases is called the $F$-matrix. It is analogous to the 6$j$ symbols encountered in the couplings of three spin-$1/2$ particles. For arbitrary numbers of anyons, different sequences of fusion-basis transformations, starting from a particular fusion-basis ending in another one, must be equal. This imposes a consistency condition on $F$-matrices known as a \textit{pentagon} equation \cite{Kitaev2006}.

On the other hand, particle exchange operators $\tau$ are represented by $R$-matrices on this fusion space. Nontrivial relations arise when we consider particle exchange operators together with $F$-matrices. These relations can be expressed by the \textit{hexagon} equations \cite{Kitaev2006}. Solving the pentagon and hexagon equations specify the consistent anyon models. For the Ising model, there are consistent solutions as follows
\begin{equation*}
R^{\epsilon \epsilon }_{1}=-1,\textrm{ \ \ \ }R^{\sigma \epsilon }_{\sigma}=R^{\epsilon\sigma }_{\sigma}=i\alpha,\textrm{ \ \ \ } R^{\sigma \sigma }_{1}=\theta
e^{i\alpha \pi /4},\textrm{ \ \ \ }R^{\sigma \sigma }_{\epsilon }=\theta
e^{-i\alpha \pi /4}
\end{equation*}%
with the possible combinations of topological spin $\theta $ (associated with the counterclockwise rotation of a particle by angle $2\pi$ around itself) and $\alpha $
\begin{equation*}
    \theta=e^{i\pi \nu/8}, \ \ \alpha=(-1)^{(\nu+1)/2}
\end{equation*}
where $\nu$ is the spectral Chern number taking odd integer values \cite{Kitaev2006}. For translationally invariant configurations of the Kitaev Honeycomb model (e.g. the reference $X,Y$-configuration i.e. $X_{q}=1$ and $Y_{q}=1$ for all $q$), the spectral Chern number is equal to $\pm1$ (depending on the direction of the magnetic field) \cite{Kitaev2006}.

The solution of the pentagon equation also determines the transformation relation between the fusion bases $\{|\Omega^{ab}_{\footnotesize\textbf{1}}\rangle,|\Omega^{ab}_{\epsilon}\rangle\}$ and $\{|\Omega^{bc}_{\footnotesize\textbf{1}}\rangle,|\Omega^{bc}_{\epsilon}\rangle\}$ of $\sigma$-particles  ($a,b, c$ and $d$) as
\begin{equation*}
    |\Omega^{ab}_{\footnotesize\textbf{1}}\rangle=\frac{|\Omega^{bc}_{\footnotesize\textbf{1}}\rangle +e^{i\varphi} |\Omega^{bc}_{\epsilon}\rangle }{\sqrt{2}}, \ \
|\Omega^{ab}_{\epsilon}\rangle =\frac{|\Omega^{bc}_{\footnotesize\textbf{1}}\rangle -e^{i\varphi}|\Omega^{bc}_{\epsilon}\rangle}{\sqrt{2}}
\end{equation*}
up to an arbitrary relative phase $e^{i\varphi}$.

In that regard, the representation $\mathcal{R}$ of the braiding operator which exchanges $\sigma$-particles $b$ and $c$ is diagonal for the fusion basis $\{|\Omega^{bc}_{\footnotesize\textbf{1}}\rangle,|\Omega^{bc}_{\epsilon}\rangle\}$
\begin{equation}
    \mathcal{R}_{\{bc\}}=\left(
      \begin{array}{cc}
        R^{\sigma \sigma }_{1} &  0 \\
        0  & R^{\sigma \sigma }_{\epsilon } \\
      \end{array}
    \right),
\end{equation}
where $\{bc\}$ stands for the basis $\{|\Omega^{bc}_{\footnotesize\textbf{1}}\rangle,|\Omega^{bc}_{\epsilon}\rangle\}$ and by using the transformation relations above, $\mathcal{R}_{\{ab\}}$ reads
\begin{equation}
    \mathcal{R}_{\{ab\}}=\frac{1}{2}\left(
      \begin{array}{cc}
        R^{\sigma \sigma }_{1}+R^{\sigma \sigma }_{\epsilon} &e^{-i\varphi}(R^{\sigma \sigma }_{1}-R^{\sigma \sigma }_{\epsilon})  \\
        e^{i\varphi}(R^{\sigma \sigma }_{1}-R^{\sigma \sigma }_{\epsilon}) & R^{\sigma \sigma }_{1}+R^{\sigma \sigma }_{\epsilon} \\
      \end{array}
    \right).\label{BerryR}
\end{equation}%

On the other hand, the Berry matrix $\mathcal{B}(\mathbf{C})$ given by Equation \eref{Berry_Kitaev1} 
is written in a different basis. This is  the same basis as $\mathcal{A}(\lambda_{1})$, where $\lambda_{1}$ is the starting point of the trajectory $\mathbf{C}$ in the parameter space. These basis states consist of two nearly degenerate energy eigenstates $\left\{ |\Phi ^{1}(\lambda_{1})\rangle ,|\Phi ^{2}(\lambda_{1})\rangle \right\}$ (see Sec.  \ref{Sec_Adia} also). We mention for completeness  that the coupling coefficient of the starting configuration is slightly different than that of the original configuration shown in Figure \ref{GConf}. We discuss the details of the trajectory used in the calculations in \nameref{Adia_Smooth}. 

Because the Berry matrix is not invariant under the basis transformation and the relation between the fusion channels and the basis $\left\{ |\Phi ^{1}(\lambda_{1})\rangle ,|\Phi ^{2}(\lambda_{1})\rangle \right\}$ is unknown, the comparison of the eigenvalues of the Berry matrix with $\left\{ R^{\sigma \sigma }_{1}, R^{\sigma \sigma }_{\epsilon} \right\}$ is more meaningful. However, before we do that we would like to point out an interesting similarity between the calculated Berry matrices and an $\mathcal{R}$-matrix: by making an analogy between four $\sigma$'s ($a,b, c$ and $d$) and four vortices (\textsf{\textbf{A}},\textsf{\textbf{B}},\textsf{\textbf{C}} and \textsf{\textbf{D}}) in Figure \ref{GConf}, for the values of $\kappa$ from the set (ii), the numerical Berry matrix $\mathcal{B}(\mathbf{C})$ converges to $\mathcal{R}_{\{\tiny \textsf{\textbf{A}}\textsf{\textbf{B}}\normalsize \}}$ with spectral Chern number $\nu =1$, namely,
\begin{equation*}
\mathcal{R}_{\{\tiny \textsf{\textbf{A}}\textsf{\textbf{B}}\normalsize \}}=\frac{1}{\sqrt{2}}%
\left(\begin{array}{cc}\label{RAB}
e^{i\pi /8} & e^{i\pi /8} \\
e^{-7i\pi /8} & e^{i\pi /8}%
\end{array}\right)
\end{equation*}
as the minimum distances $d$ increases. We point out that the arbitrary relative phase $e^{i\varphi}=-i$ is chosen for $\mathcal{R}_{\{\tiny \textsf{\textbf{A}}\textsf{\textbf{B}}\normalsize \}}$; and a similar arbitrary relative phase is fixed between the basis states $|\Phi ^{1}(\lambda_{1})\rangle$ and $|\Phi ^{2}(\lambda_{1})\rangle$. This arbitrary phase reflects a gauge freedom and is chosen to obtain the best agreement between $\mathcal{B}(\mathbf{C})$ to $\mathcal{R}_{\{\tiny \textsf{\textbf{A}}\textsf{\textbf{B}}\normalsize \}}$. 

\begin{figure}[t]
\begin{flushright}
\includegraphics[width=400pt]{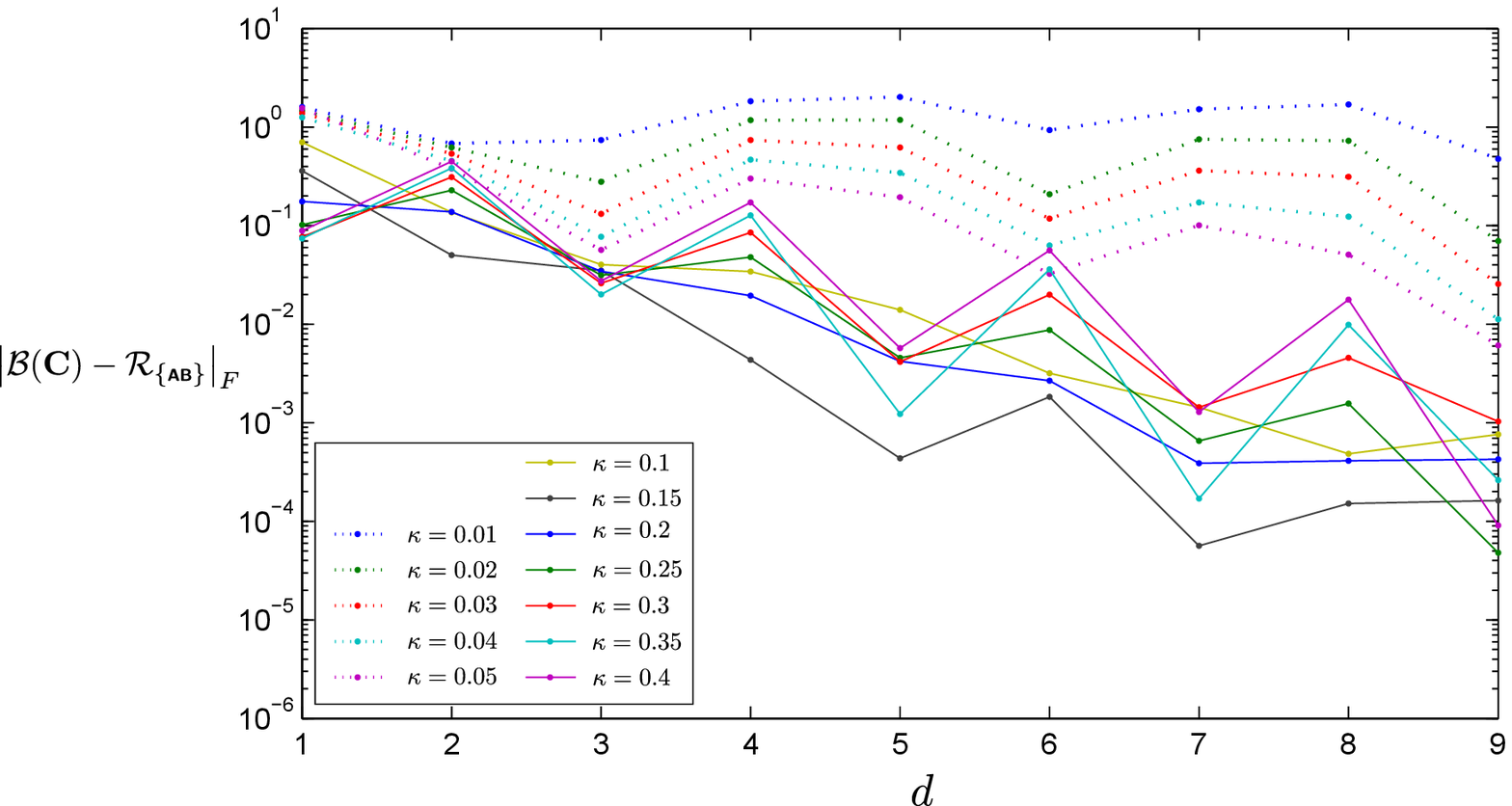}\\[0pt]
\end{flushright}
\caption{(color online) $\left|\mathcal{B}(\mathbf{C})-\mathcal{R}_{\{\tiny \textsf{\textbf{A}}\textsf{\textbf{B}}\normalsize \}}\right|_{F}$ : Frobenius distance between $\mathcal{B}(\mathbf{C})$ and $\mathcal{R}_{\{\tiny \textsf{\textbf{A}}\textsf{\textbf{B}}\normalsize \}}$ vs. $d$. Note that the numerical details of the calculations are given at the beginning of the Section \ref{Sec_Adia}.}
\label{FrobB}
\end{figure}

The numerical results are summarized in Figure \ref{FrobB} where the Frobenius norm was used to measure the distance between the two matrices. 
The Frobenius norm of an $n\times n$ matrix $A$ is defined as $ |A|_{F}=\left(\sum_{i,j}|A_{ik}|^{2}\right)^{1/2}=\sqrt{\textrm{tr}(AA^{\dagger})}.$
Here, we calculate the norm $|\mathcal{R}_{\{\tiny \textsf{\textbf{A}}\textsf{\textbf{B}}\normalsize \}}-\mathcal{B}(\mathbf{C})|_{F}$ to measure the distance between $\mathcal{B}(\mathbf{C})$ and $\mathcal{R}_{\{\tiny \textsf{\textbf{A}}\textsf{\textbf{B}}\normalsize \}}$. The high level precision of the results is reflected in the unitarity of the calculated Berry matrix: $\left|\mathcal{B}(\mathbf{C})\mathcal{B}^{\dagger}(\mathbf{C})-I\right|_{F} < 10^{-5}$ for all $d$ and $\kappa$. It is not difficult to show that the maximum Frobenius distance between two unitary 2$\times$2 matrices is equal to 2, so to evaluate the error of the calculations the Frobenius distance has to be divided by 2.

As an example, for the vortex configuration illustrated in Figure \ref{GConf} with $d=9$ and $\kappa =0.25$, the numerical value of the Berry matrix $\mathcal{B}(\mathbf{C})$ is
\begin{equation*}
\mathcal{B}(\mathbf{C})=%
\left(\begin{array}{cc}
0.653270 + 0.270630i & 0.653280+0.270598i \\
-0.653280-0.270598i & 0.653296+0.270568i%
\end{array}\right)%
\end{equation*}%
whose Frobenius distance to $\mathcal{R}_{\{\tiny \textsf{\textbf{A}}\textsf{\textbf{B}}\normalsize \}}$ given in Equation  \eref{RAB}  
is equal to $4.8\times 10^{-5}$ and the normalized error associated with  $\mathcal{B}(\mathbf{C})$ compared to the exact result from the effective field theory $\mathcal{R}_{\{\tiny \textsf{\textbf{A}}\textsf{\textbf{B}}\normalsize \}}$ is 0.0024\%.

This similarity between the basis $\{|\Phi ^{1}(\lambda_{1})\rangle, |\Phi ^{2}(\lambda_{1})\rangle\}$ and  $\{|\Omega^{\tiny \textsf{\textbf{A}}\textsf{\textbf{B}}\normalsize }_{I} \rangle, |\Omega^{\tiny \textsf{\textbf{A}}\textsf{\textbf{B}}\normalsize}_{\epsilon}\rangle\}$
 can be understood from the perspective of Majorana fermions. First of all, note that the eigenstates $|\Phi ^{1}\rangle$ and $|\Phi ^{2}\rangle$ are also the eigenstates of the operators $i \gamma_{1a} \gamma_{1b}$ and $i \gamma_{2a} \gamma_{2b}$ where \begin{eqnarray*}
  &&\gamma_{1a}=\beta^{\dag}_{1}+\beta_{1}, \textrm{\ \ \ \ } \gamma_{1b}=i(\beta^{\dag}_{1}-\beta_{1}), \\
  &&\gamma_{2a}=\beta^{\dag}_{2}+\beta_{2}, \textrm{\ \ \ \ } \gamma_{2b}=i(\beta^{\dag}_{2}-\beta_{2}),
\end{eqnarray*}
are Majorana fermions, which are their own anti-particles. Kitaev showed that for odd values of the Chern number vortices carry Majorana fermions \cite{Kitaev2006}. The localization of the Majorana fermions around vortices are also demonstrated numerically in \cite{Kells2010} by expressing them in terms of $c$-fermions. In that respect, the equality of the bases can be understood as the Majorana fermion pairs $(\gamma_{1a}, \gamma_{1b})$ and $(\gamma_{2a}, \gamma_{2b})$ being localized on the vortex pairs $(\textsf{\textbf{A}},\textsf{\textbf{B}})$ and $(\textsf{\textbf{C}},\textsf{\textbf{D}})$. Note that because the localization of Majorana fermions is only related to the eigenstates of the Hamiltonian, it does \textit{not} depend on the history of the vortex motion. That is, the configuration properties (such as relative distance of vortices to each other, boundary conditions, the values of $L_{x}$ and $L_{y}$ etc.) determine the relation between the energy basis and the fusion basis. For example, for the same size configurations with $L_{x}=1$ and $L_{y}=-1$, the Berry phase $\mathcal{B}(\mathbf{C})$ is diagonal so that $\{|\Phi ^{1}(\lambda_{1})\rangle, |\Phi ^{2}(\lambda_{1})\rangle\} \cong \{|\Omega^{\tiny \textsf{\textbf{B}}\textsf{\textbf{C}}\normalsize }_{I} \rangle, |\Omega^{\tiny \textsf{\textbf{B}}\textsf{\textbf{C}}\normalsize}_{\epsilon}\rangle\}$. (Berry phase calculations for the other $L_{x}, L_{y}$ and $N_{x}, N_{y}$ configurations will be presented in a different context in future.)

\begin{figure}[t]
\begin{flushright}
\includegraphics[width=440pt]{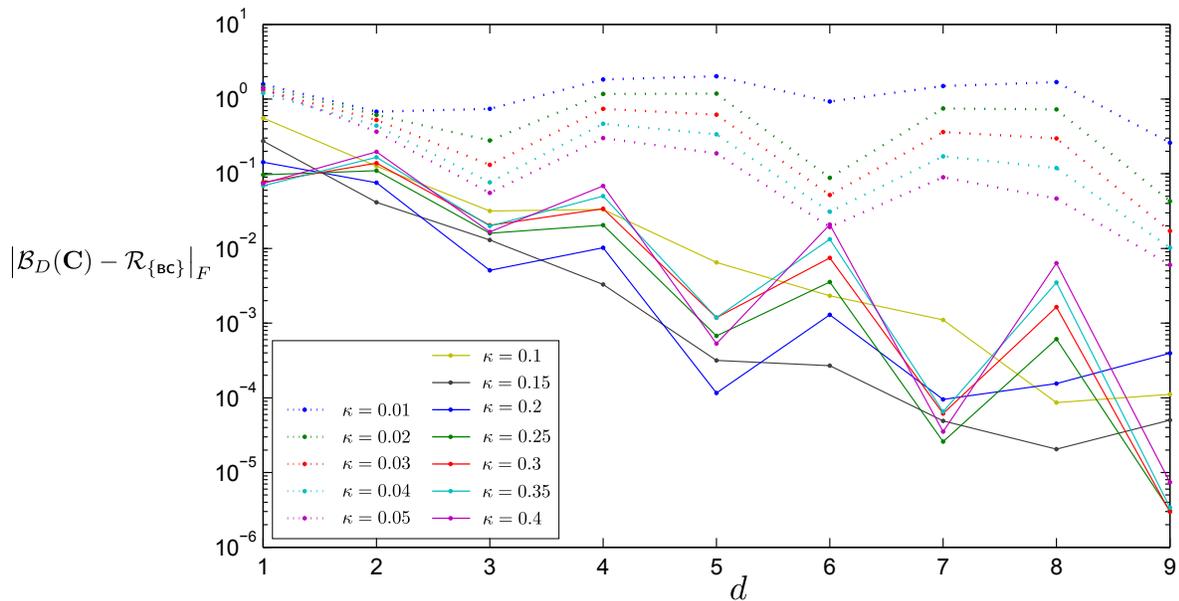}\\[0pt]
\end{flushright}
\caption{(color online) $\left|\mathcal{B}_{D}(\mathbf{C})-\mathcal{R}_{\{\tiny \textsf{\textbf{B}}\textsf{\textbf{C}}\normalsize \}}\right|_{F}$ : Frobenius distance between diagonalized $\mathcal{B}_{D}(\mathbf{C})$ and $\mathcal{R}_{\{\tiny \textsf{\textbf{B}}\textsf{\textbf{C}}\normalsize \}}$ vs. $d$.  Note that the numerical details of the calculations are given at the beginning of the Section \ref{Sec_Adia}.}
\label{Frob3991}
\end{figure}

On the other hand, the eigenvalue comparison is presented in Figure \ref{Frob3991} where $\mathcal{B}_{D}(\mathbf{C})$ is the diagonalized form of $\mathcal{B}(\mathbf{C})$. To emphasize the proximity of the eigenvalues of $\mathcal{B}(\mathbf{C})$ (l.h.s.) to those of $\mathcal{R}_{\{\tiny \textsf{\textbf{A}}\textsf{\textbf{B}}\normalsize \}}$ (r.h.s.), the explicit values for $d=9$ and $\kappa =0.30$ are 
\begin{eqnarray*}
0.3826813+0.9238804i &\simeq & e^{3i\pi /8} \\
0.9238802-0.3826817i &\simeq & e^{-i\pi /8}
\end{eqnarray*}%
giving the Frobenius distance $3\times 10^{-6}$ and the error 0.00015\%.

The exponential convergence of the Berry phase to the expected values as the minimum distance $d$ between the vortices increases, which the accuracy of our calculations allows us to observe, reveals subtle connections. The oscillations and the agreement of the calculated Berry phase to the expected values are in correlation with the exact ground state degeneracy and the gapfulness of the system (Figure \ref{Gap}) with few exceptions. However, one should keep in mind that although the ground state degeneracy and the gapfulness of the system play a role in the physical evolution of the system, the Berry phase only depends on the eigenstates. Therefore, all the Berry phase values should be understood in terms of the properties of the eigenstates.

At this point, it is essential to note that the structure of the eigenstates determines the spread of the localization of the Majorana fermions around the vortices \cite{Kells2010}. Because the braiding properties of the non-Abelian anyons are determined when the particles are away from each other to avoid tunneling, the radius of the Majorana fermions around the vortices affects the calculated Berry matrices. Thus, the relation between the calculated Berry phase to the expected $\mathcal{R}$ values is also an indication of how tightly the Majorana fermions are localized around the vortices.

In that respect, the correlations between Figure \ref{Gap} and Figure \ref{Frob3991} suggests that the bigger gaps and the nearer degenerate states make the localization of the Majorana fermions tighter around the vortices. However, it seems that this is not always the case as is seen for the systems with $d=6$ and the small values of $\kappa$ (i.e. from the set (i)).

\section{Outlook and Summary}

To summarize, we have analyzed some four-vortex configurations of the Kitaev honeycomb model (see Figure \ref{GConf}) under various magnetic fields. All lattice configurations are even-by-even in terms of the number of plaquettes and they are defined on a torus whose size is determined by the minimum distance $d$ between the vortices. As was shown in Section \ref{Sec_Adia}, the ground states of such configurations are two-fold nearly degenerate. Under the adiabatic interchange of the vortices, the evolution of the system is restricted to the ground states and it is governed by the Berry matrix whose numerical evaluation was presented in detail in Section \ref{Sec_Num}. The numerical results show that as $d$ increases the small gap between nearly degenerate state decreases exponentially with some oscillations (see Figure \ref{Gap}) also the calculated Berry matrices get exponentially closer to the expected ones with similar oscillations (see Figure \ref{Frob3991}).

The main result of the paper is the explicit demonstration of the non-Abelian statistics by numerically calculating the Berry matrix which governs the evolution of the system under the adiabatic interchange of vortices of the Kitaev honeycomb model. We showed that the resulting Berry matrices exponentially converge to the expected braiding properties of Ising anyons derived from the effective field theory. The presented method for the Berry phase calculations represents a direct approach to the non-Abelian statistics, and it can be used to study braiding properties of anyons for any configurations.
The high accuracy of the method also allows us to test the stability of the Berry phase, so that we may make conclusions about both its implementation under (more) realistic conditions and the pitfalls we should expect in such situations. Moreover, the accuracy of the calculations presented here can provide important predictive power in the analysis of other, less understood, topological phases. They also reveal the dependence on details of the implementation of the braiding operations and thus allow testing of non-adiabatic and other effects which will likely constitute a dominant source of errors in topological quantum information processing. The accuracy of the calculations shows the exact dependence of the simulated Berry phase on the details of the model, like the splitting of the ground state levels intrinsic to any finite system. Possible applications thus extend to modeling and implementation of  quantum information protocols whose reliability can be tested under various effects; for example, disorder. Naturally, the first step in this potentially fruitful story is the demonstration of highly accurate and sufficiently large scale direct simulation of the Berry phase as we have presented in this manuscript. Moreover, the method is general enough to be useful in the study of non-Abelian statistics in other models including the Yao-Kivelson model \cite{Yao2007}, square-octagon model \cite{Kells2010c} or any system with a quadratic fermionic Hamiltonian.

\ack
We would like to thank Graham Kells, Steve Simon and Paul Watts for discussion and valuable comments, and to Niall Moran for his assistance with high performance computing. This work has been supported by Science Foundation Ireland (SFI) through the Awards 05/YI2/I680 and 10/IN.1/I3013. The authors acknowledge the SFI/HEA Irish Centre for High-End Computing (ICHEC) for the provision of computational facility and support.

\appendix
\section*{Appendix A.}\label{Adia_Smooth}
\setcounter{section}{1}

When we change the parameters of the Hamiltonian so as to follow a smooth closed curve in the parameter space, the evolution of the system is governed by the Berry phase if the process is adiabatic \cite{Bohm2009}. In this Appendix, we will introduce the trajectory which we use and then discuss the adiabaticity of the exchange process in detail.

\begin{figure}[t]
\begin{center}
\includegraphics[width=350pt]{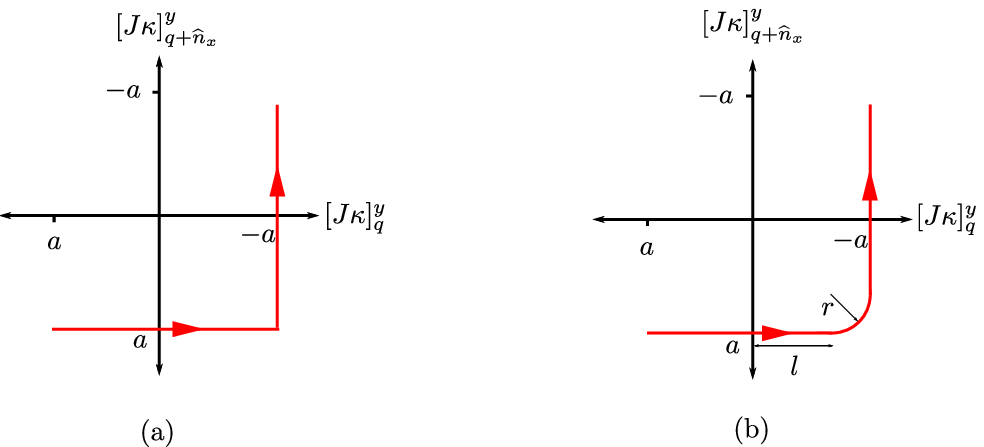}\\[0pt]
\end{center}
\caption{(Color online) (a) Changing the $J$ and $\kappa $ values of a link
after another gives us a trajectory with square edges.  (b) Changing the $J$ and $\kappa $ values of the next link before stopping to change that of the present one gives us a smooth trajectory.}
\label{circular}
\end{figure}

To examine the smoothness of the trajectory shown in Figure \ref{GConf}, let us take as an example the $[J\kappa]$ of the first two links of the path, namely $[J\kappa]_{q}^{y}$ and $[J\kappa]_{q+\widehat{n}_{x}}^{y}$ where $q=(2,N_{y})$. We need to change $[J\kappa]_{q}^{y}$ from $a$ to $-a$ to move vortex \textbf{B} to the right plaquette where $a=1$ for $J$, and $a=l\times0.01$ or $a=k\times0.05$ for $\kappa$ where $l=1,...,5$, $k=2,...,8$. Changing the sign of $[J\kappa]_{q}^{y}$ first, and then $[J\kappa]_{q+\widehat{n}_{x}}^{y}$, results in a trajectory with discontinuous edges as shown in Figure \ref{circular}(a). To move the vortex smoothly, we need to start changing $[J\kappa]_{q+\widehat{n}_{x}}^{y}$ before $[J\kappa]^{y}_{q}$ reaches $-a$, as shown in Figure \ref{circular}(b). For this reason a vortex is never completely localized on a particular plaquette. We will refer to the part where only one $[J\kappa]$ changes as the \textit{linear part} and the part where two $[J\kappa]$ changes as the \textit{circular part}. In our numerical calculations, every link on the trajectory is sampled with 4000 data points (3991 linear plus 9 circular) which are separated by an \textit{equal distance} $\Delta s$ in the parameter space. Let $l$ be the length of the linear part from 0 to the beginning of the circular part (see Figure \ref{circular}(b)) and $r$ the radius of the circular part; then $r+l=|a|$ and $\Delta s= \Delta l=r\Delta \theta$ where $\Delta l= l/3991 $ and $\Delta\theta=\frac{\pi}{2}/9$, so that $l=0.997|a|$ and $r=0.003 |a|$.

The starting point of the trajectory (i.e. $\lambda_{1}$) is chosen to be the beginning of the linear part of the $[J\kappa]$ of the first link (i.e. $[J\kappa]_{(2,N_{y})}^{y}$) of the path; because of that the coupling coefficients of the starting configuration are slightly different than that of the original configuration shown in Figure \ref{GConf}.

On the other hand, the first requirement for a process to be adiabatic is the gap between the eigenspace and the rest of the Hamiltonian's spectrum and this has been analyzed in Figure \ref{Gap}. The standard condition for maintaining the adiabaticity for the $n^{th}$ eigenstate is as follows:
\begin{equation*}
   \sum_{m\neq n} \left|\frac{\langle\Phi^{n}|\dot{\Phi}^{m}\rangle}{E^{n}-E^{m}}\right|\ll1,
\end{equation*}
where dot denotes the time derivative. However, this condition is only valid if $\langle\Phi^{n}(\lambda)|\dot{\Phi}^{m}(\lambda)\rangle$ and $E^{m}(\lambda)-E^{n}(\lambda)$ are constant for all $m$ through the trajectory \cite{Tong2007}. When $\langle\Phi^{n}(\lambda(t))|\dot{\Phi}^{m}(\lambda(t))\rangle$ and $E^{m}(\lambda)-E^{n}(\lambda)$ are not constant the validity condition of the adiabatic approximation can be written as
\begin{equation*}
   v\frac{\max\left(\sum_{m\neq n}|\langle \frac{d }{d\lambda}\Phi^{n}(\lambda)|\Phi^{m}(\lambda)\rangle|\right)}{\min|E^{n}(\lambda)-E^{m}(\lambda)|}\ll1.
\end{equation*}
where $v= \frac{d \lambda}{dt}$ is the speed of the process and we used the following equality
\begin{equation*}
\langle\Phi^{n}(\lambda(t))|\dot{\Phi}^{m}(\lambda(t))\rangle =-\langle\dot{\Phi}^{n}(\lambda(t))|\Phi^{m}(\lambda(t))\rangle
\end{equation*}
which can be verified by differentiating the orthogonality condition $\langle\Phi^{n}(\lambda)|\Phi^{m}(\lambda)\rangle=0$.

First we define 
\begin{equation}
\left|\frac{d}{d\lambda}\Phi^{n_{i}}(\lambda)\right|_{\bot}:= \left|\frac{d}{d\lambda}\Phi^{n_{i}}(\lambda)\right|-\sum^{k}_{j=1}|\langle\frac{d }{d\lambda}\Phi^{n_{i}}(\lambda)|\Phi^{n_{j}}(\lambda)\rangle|
\end{equation}
for a $k$-fold degenerate eigenspace with basis $|\Phi^{n_{i}}(\lambda)\rangle$ for $i=1,...,k$. Then by using the following inequality
\begin{equation*}
\sum_{m\neq n}\left|\langle\frac{d }{d\lambda}\Phi^{n}(\lambda)|\Phi^{m}(\lambda)\rangle\right| <\left|\frac{d }{d\lambda}\Phi^{n}(\lambda)\right|_{\bot},
\end{equation*}
we can rewrite the general validity condition for adiabaticity as
\begin{equation*}
   v\frac{\max\left|\frac{d}{d\lambda}\Phi^{n_{i}}(\lambda)\right|_{\bot}} {\min|E^{n}(\lambda)-E^{m}(\lambda)|}\ll1.
\end{equation*}

\begin{figure}[t]
\begin{center}
\includegraphics[width=440pt]{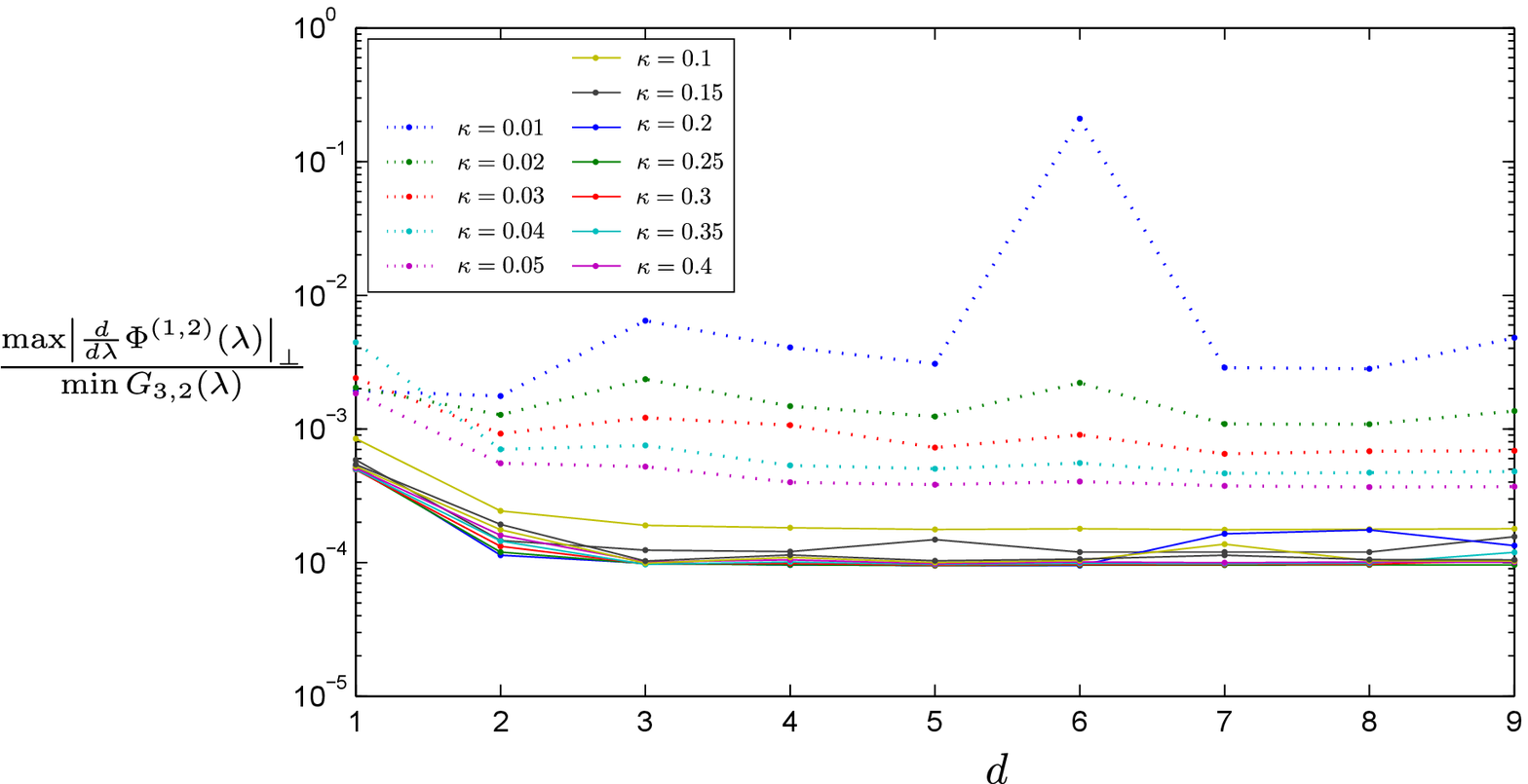}\\[0pt]
\end{center}
\caption{(Color online) Adiabaticity condition $\frac{\max\left|\frac{d}{d\lambda}\Phi^{(1,2)}(\lambda)\right|_{\bot}}{\min G_{3,2}(\lambda)}$ where $\max\left|\frac{d}{d\lambda}\Phi^{(1,2)}(\lambda)\right|_{\bot}:=\max\left(\left|\frac{d}{d\lambda}\Phi^{1}(\lambda)\right|_{\bot}, \left|\frac{d}{d\lambda}\Phi^{2}(\lambda)\right|_{\bot}\right)$ vs. $d$.  The numerical details of the calculations are given at the beginning of the Section \ref{Sec_Adia}.}
\label{Adia}
\end{figure}

Figure \ref{Adia} shows the validity of the adiabatic approximation for various $\kappa$ values against $d$. The smaller the values of $\kappa$, the smaller the gap to the first excited state, and thus the smaller the speed $v$ is required to maintain adiabaticity.

\appendix
\section*{Appendix B.}\label{BM}
\setcounter{section}{2}

Here, we would like to show by using the Bloch-Messiah theorem \cite{Ring2004} that the absolute value of the overlap $|\langle \phi _{0}|\phi _{k}\rangle|$ between the ground states of $\beta(0)$ and $\beta(k)$-fermions can be calculated as $|\langle \phi _{0}|\phi _{k}\rangle| =\sqrt{|\det U(k,0)| }$. Let us start from Equation \eref{gammai0}:
\begin{equation}
\left[\begin{array}{cc}
\beta _{\leftrightarrow }^{\dag }(k) & \beta _{\leftrightarrow }(k)\\
 \end{array} \right]=\left[\begin{array}{cc}
\beta _{\leftrightarrow }^{\dag }(0) & \beta _{\leftrightarrow }(0)\\
 \end{array} \right]T(k,0)
\end{equation}
where
\begin{equation*}
T(k,0):=T^{\dag}(0)T(k)=\left[\begin{array}{cc}
U(k,0) & V^{\ast }(k,0) \\
V(k,0) & U^{\ast }(k,0) \\
 \end{array} \right]
\end{equation*}
and
\begin{eqnarray*}
U(k,0) &:=&U^{\dag }(0)\ U(k)+V^{\dag }(0)\ V(k)  \\
V(k,0) &:=&V^{T}(0)\ U(k)+U^{T}(0)\ V(k).
\end{eqnarray*}

The Bloch-Messiah decomposition of $T(k,0)$ reads
\begin{equation*}
\fl T(k,0)=\left[\begin{array}{cc}
D(k,0) & 0 \\
0 & D^{\ast }(k,0)\\
 \end{array} \right]
\left[\begin{array}{cc}
\overline{U}(k,0) & \overline{V}(k,0) \\
\overline{V}(k,0) & \overline{U}(k,0)\\
 \end{array} \right]
\left[\begin{array}{cc}
C(k,0) & 0 \\
0 & C^{\ast }(k,0)\\
 \end{array} \right].
\end{equation*}%

Let us define $\overline{\beta }(0)$-fermions as%
\begin{equation*}
\overline{\beta }_{k}^{\dag }(0)=\sum_{k^{\prime }}D_{k^{\prime
}k}(k,0)\beta _{k^{\prime }}^{\dag }(0)
\end{equation*}%
Now, we can write the vacuum $|$$\phi _{(k,0)}\rangle $ of $\beta
(k)$-fermions in terms of the vacuum $|$$\phi _{0}\rangle $ of $\beta
(0)$-fermions as
\begin{equation}
|\phi _{(k,0)}\rangle =\prod\limits_{i}\overline{\beta }_{i}^{\dag
}(0)\,\prod\limits_{p}\left(u_{p}(k,0)+v_{p}(k,0)\:\overline{\beta }_{p}^{\dag }(0)%
\:\overline{\beta }_{\overline{p}}^{\dag }(0)\right)\,|\phi _{0}\rangle.
\label{BM onishi}
\end{equation}
Note that $|$$\phi _{k}\rangle $ and $|$$\phi _{(k,0)}\rangle $ are the same up to an overall phase.

Now the overlap $\langle \phi _{0}|\phi _{(k,0)}\rangle $ reads%
\begin{equation*}
\langle \phi _{0}|\phi _{(k,0)}\rangle =\langle \phi _{0}|\,\prod\limits_{i}%
\overline{\beta }_{i}^{\dag }\,\prod\limits_{p}\left(u_{p}(k,0)+v_{p}(k,0)%
\:\overline{\beta }_{p}^{\dag }(0)\:\overline{\beta }_{\overline{p}}^{\dag
}(0)\right)\,|\phi _{0}\rangle.
\end{equation*}%
This is only non-zero if there
is no product with index $i$ and it is equal to
\begin{equation*}
\fl \langle \phi _{0}|\phi _{(k,0)}\rangle =\langle \phi
_{0}|\prod\limits_{p}u_{p}(k,0)\textrm{ }|\phi _{0}\rangle
=\prod\limits_{p}u_{p}(k,0)=\sqrt{\det \overline{U}(k,0)}.
\end{equation*}
Recall that $\overline{U}(k,0)$ is diagonal with positive entries. Because of the arbitrariness of the phase of Equation \eref{BM onishi}, the overlap $\langle \phi _{0}|\phi _{(k,0)}\rangle$ can be taken to be positive; hence $|\langle \phi _{0}|\phi _{k}\rangle| = \langle \phi _{0}|\phi _{(k,0)}\rangle$.

Moreover, since $U(k,0)=D(k,0) \overline{U}(k,0) C(k,0)$, and $D(k,0)$ and $D(k,0)$ are unitary, $\sqrt{\det \overline{U}(k,0)}=\sqrt{|\det U(k,0)| }$.
Therefore,
\begin{equation*}
\langle \phi _{0}|\phi _{k}\rangle =\sqrt{|
\det U(k,0)| }.
\end{equation*}

\appendix
\section*{Appendix C.}\label{Pfaffian}
\setcounter{section}{3}

Here, we would like to show that the overlap
\begin{equation*}
\fl \langle \psi _{(l,0)}|\psi _{(k,0)}\rangle =\sqrt{|\det U(l,0)|}\textrm{ }%
\sqrt{|\det U(k,0)|}\textrm{ \ \ } \left( -1\right) ^{N\left( N+1\right) /2}~\textrm{Pf}(\mathcal{Z}%
(l,0;k,0))
\end{equation*}%
where 
\begin{eqnarray}
\mathcal{Z}(l,0;k,0)=
\left[\begin{array}{cc}
Z(k,0) & -I \\
I & -Z^{\ast }(l,0) \\
 \end{array} \right],
 \end{eqnarray} and $N$ is the number of fermions and therefore also the size of $Z$ matrices, can be written as
\begin{eqnarray*}
\langle \psi _{(l,0)}|\psi _{(k,0)}\rangle = \left( -1\right) ^{N\left( N+1\right) /2}\sqrt{%
\exp \left\{ i\theta _{0}(l,k) \right\} ~|\det U(l,k) |}
\end{eqnarray*}%
 where
\begin{eqnarray*}
U(l,k) &:=&U^{\dag }(k)U(l)+V^{\dag }(k)V(l) \\
\theta _{0}(l,k) &=&\arg \left\{ \det U(k,0)\det U^{\dag
}(l,0)\det U(l,k) \right\}.
\end{eqnarray*}

First of all note that for any skew-symmetric matrix $K$, $\textrm{Pf}(K)=\sqrt{\textrm{det}(K)}$. Moreover, if
$K= \left[\begin{array}{cc}
A & B \\
C & D \\
\end{array} \right]$
where $A,B,C,D$ are $n\times n$ matrices with complex coefficients and $%
CD=DC,$ then $\det K=\det \left( AD-BC\right)$ \cite{Silvester2000}. Therefore,
\begin{equation}
\textrm{Pf}(\mathcal{Z}(l,0;k,0))= \sqrt{\det\mathcal{Z}(l,0;k,0)}=\sqrt{\det \left( I-Z(k,0)Z^{\ast }(l,0)\right)}. \label{B1}
\end{equation}

Recall that $Z(k,0)=\left( V(k,0)U^{-1}(k,0)\right) ^{\ast }$. By using the conditions that $U(k,0)$ and $V(k,0)$ need to satisfy in order for the matrix $T(k,0)$ to be unitary,
%\begin{equation*}
%T(k,0)T(k,0)^{\dag}=T(k,0)^{\dag}T(k,0)=I,
%\end{equation*}
it can be shown that $Z(k,0)$ is also skew-symmetric and can be written as 
\begin{equation*}
 Z(k,0)=-\left( U^{\dag }(k,0)\right)
^{-1}V^{\dag }(k,0).
\end{equation*}
Therefore, Equation \eref{B1} reads
\begin{eqnarray*}
\fl I-Z(k,0)Z^{\ast }(l,0) &=&I+(U^{\dag }(k,0))^{-1 }V^{\dag }(k,0)V(l,0)U^{-1}(l,0) \\
&=&(U^{\dag }(k,0))^{-1 }\left( U^{\dag }(k,0)U(l,0)+V^{\dag }(k,0)V(l,0)\right)
U^{-1}(l,0).
\end{eqnarray*}%
By using Equation \eref{UV definitions}, it is easy to show
\begin{equation*}
U^{\dag }(k,0)U(l,0)+V^{\dag }(k,0)V(l,0)=U\left( l,k\right),
\end{equation*}%
where $U(l,k) :=U^{\dag }(k)U(l)+V^{\dag }(k)V(l)$ is a generalized version of Equation \eref{UV definitions}.

Therefore we have%
\begin{equation*}
\det \left( I-Z(k,0)Z^{\ast }(l,0)\right) =\det \left( (U^{\dag
}(k,0))^{-1} U(l,k) U^{-1}(l,0)\right).
\end{equation*}%
Finally, we can express $\langle \phi _{0}|e^{\mathbf{Z}^{\dag }(l,0)}e^{%
\mathbf{Z}(k,0)}|\phi _{0}\rangle $ as
\begin{equation*}
\langle \phi _{0}|e^{\mathbf{Z}^{\dag }(l,0)}e^{\mathbf{Z}(k,0)}|\phi
_{0}\rangle =\left( -1\right) ^{N\left( N+1\right) /2}\sqrt{\frac{\det U(l,k) }{\det U^{\dag }(k,0)\det U(l,0)%
}},
\end{equation*}
and express the overlap between $|$$\psi _{(k,0)}\rangle $ and $|$$\psi _{(l,0)}\rangle $ as
\begin{eqnarray*}
\fl \langle \psi _{(l,0)}|\psi _{(k,0)}\rangle &=&\sqrt{|\det U(l,0)|}\textrm{\ }\sqrt{|\det U(k,0)|}\textrm{ \ }\left(
-1\right) ^{N\left( N+1\right) /2}\sqrt{\frac{\det U(l,k) }{\det U^{\dag }(k,0)\det U(l,0)}} \\
\fl &=&\left( -1\right) ^{N\left( N+1\right) /2}\sqrt{%
\exp \left\{ i\theta _{0}(l,k) \right\} ~|\det U\left(
j,i\right) |},
\end{eqnarray*}%
 where
\begin{equation*}
\theta _{0}(l,k) =\arg \left\{ \det U(k,0)\det U^{\dag
}(l,0)\det U(l,k) \right\}.
\end{equation*}%

\section*{References}
\bibliographystyle{unsrt}

\end{document}